\documentclass[journal=ancac3, manuscript=article, layout=twocolumn]{achemso} 
\setkeys{acs}{articletitle = true}

\usepackage[version=3]{mhchem} 
\usepackage{siunitx}    
\usepackage{pdflscape}  
\usepackage{rotating}   
\usepackage{textgreek}  
\usepackage{gensymb}    
\usepackage[misc]{ifsym} 
\usepackage{orcidlink}  
\usepackage{listings}   
\usepackage{colortbl}   
\usepackage{tabularx}   
\usepackage{longtable}  
\usepackage{subcaption}
\usepackage{multirow}
\usepackage{snotez}     
\usepackage{csquotes}   
\usepackage{tabularx}
\usepackage{placeins}
\usepackage{hep-float} 
\usepackage{tablefootnote}
\usepackage{foreign}

\usepackage{microtype}
\usepackage{xurl}

\mciteErrorOnUnknownfalse

\DeclareSIUnit\sccm{sccm}
\DeclareSIUnit\torr{Torr}

\author{Lukas Gr\"unewald}
\affiliation{Authors contributed equally to this work.}
\alsoaffiliation{EMAT, University of Antwerp, Groenenborgerlaan 171, B-2020 Antwerp, Belgium}

\author{Dmitry Chezganov}
\affiliation{Authors contributed equally to this work.}
\alsoaffiliation{EMAT, University of Antwerp, Groenenborgerlaan 171, B-2020 Antwerp, Belgium}
\alsoaffiliation{NANOlab Center of Excellence, University of Antwerp, Groenenborgerlaan 171, B-2020 Antwerp, Belgium}

\author{Robin De Meyer}
\affiliation{Authors contributed equally to this work.}
\alsoaffiliation{EMAT, University of Antwerp, Groenenborgerlaan 171, B-2020 Antwerp, Belgium}
\alsoaffiliation{NANOlab Center of Excellence, University of Antwerp, Groenenborgerlaan 171, B-2020 Antwerp, Belgium}
\alsoaffiliation{PLASMANT, University of Antwerp, Universiteitsplein 1, B-2610 Antwerpen-Wilrijk, Belgium}

\author{Andrey Orekhov}
\affiliation{EMAT, University of Antwerp, Groenenborgerlaan 171, B-2020 Antwerp, Belgium}

\author{Sandra Van Aert}
\affiliation{EMAT, University of Antwerp, Groenenborgerlaan 171, B-2020 Antwerp, Belgium}
\alsoaffiliation{NANOlab Center of Excellence, University of Antwerp, Groenenborgerlaan 171, B-2020 Antwerp, Belgium}

\author{Annemie Bogaerts}
\affiliation{PLASMANT, University of Antwerp, Universiteitsplein 1, B-2610 Antwerpen-Wilrijk, Belgium}

\author{Sara Bals}
\affiliation{EMAT, University of Antwerp, Groenenborgerlaan 171, B-2020 Antwerp, Belgium}
\alsoaffiliation{NANOlab Center of Excellence, University of Antwerp, Groenenborgerlaan 171, B-2020 Antwerp, Belgium}

\author{Jo Verbeeck}
\affiliation{EMAT, University of Antwerp, Groenenborgerlaan 171, B-2020 Antwerp, Belgium}
\email{jo.verbeeck_at_uantwerpen.be}

\title{\emph{In-situ} Plasma Studies using a Direct Current Microplasma in a Scanning Electron Microscope}


\begin{document}

\begin{abstract}
Microplasmas can be used for a wide range of technological applications and to improve our understanding of fundamental physics. 
Scanning electron microscopy, on the other hand, provides insights into the sample morphology and chemistry of materials from the \unit{\mm}- down to the \unit{\nm}-scale.
Combining both would provide direct insight into plasma-sample interactions in real-time and at high spatial resolution.
Up till now, very few attempts in this direction have been made, and significant challenges remain.
This work presents a stable direct current glow discharge microplasma setup built inside a scanning electron microscope. 
The experimental setup is capable of real-time \emph{in-situ} imaging of the sample evolution during plasma operation and it demonstrates localized sputtering and sample oxidation.
Further, the experimental parameters such as varying gas mixtures, electrode polarity, and field strength are explored and experimental $V$-$I$ curves under various conditions are provided.
These results demonstrate the capabilities of this setup in potential investigations of plasma physics, plasma-surface interactions, and materials science and its practical applications.
The presented setup shows the potential to have several technological applications, e.g., to locally modify the sample surface (e.g., local oxidation and ion implantation for nanotechnology applications) on the \unit{\um}-scale.
\end{abstract}

\section{Introduction} \label{intro}

A plasma is a complex and versatile state of matter with many established applications, e.g., in the semiconductor industry~\cite{kanarikMysteriousWorldPlasma2020}, as well as promising emerging technologies. 
The potential of plasma technology spans a broad range, including biomedical applications~\cite{laroussiLowTemperaturePlasmaJet2015}, materials science~\cite{linMicroplasmaNewGeneration2015}, and gas conversion for environmental applications~\cite{bogaertsGasDischargePlasmas2002}. 
Many different types of plasma exist~\cite{fridmanPlasmaPhysicsEngineering2011}, but the simplest geometry consists of two electrodes separated by a gas (at low, atmospheric, or elevated pressure), with a voltage being applied between the electrodes~\cite{gudmundssonFoundationsDCPlasma2017}. 
More recently, so-called \enquote{microplasmas} have received a rising interest in the scientific community~\cite{chiangMicroplasmasAdvancedMaterials2020, schoenbach20YearsMicroplasma2016, izaMicroplasmasSourcesParticle2008, foestMicroplasmasEmergingField2006, karanassiosMicroplasmasChemicalAnalysis2004}.
Microplasmas have at least one dimension in the sub-\unit{\mm} range~\cite{schoenbach20YearsMicroplasma2016}.
Besides the (i)~practical aspect of reduced operation cost of microplasma setups compared to large plasma reactors for laboratory-scale experiments and (ii)~a general trend toward miniaturization of devices in plasma-application areas, microplasmas also have interesting properties.
For example, the large surface-to-volume ratio and short gap distances between the electrodes (typically a few \SI{100}{\um}) lead to a non-equilibrium state where the ion/gas temperature is lower than the electron temperature~\cite{linMicroplasmaNewGeneration2015}. 
This results in a \enquote{cold} plasma with gas temperatures close to room temperature~\cite{izaMicroplasmasSourcesParticle2008, schoenbach20YearsMicroplasma2016, chiangMicroplasmasAdvancedMaterials2020}, which shows great promise, e.g., in nanomaterial and nanoparticle fabrication~\cite{linMicroplasmaNewGeneration2015}.
In addition, microplasmas are not confined to vacuum operation. 
Paschen's law relates the breakdown voltage of a gas with the product $pd$ of the pressure $p$ and the gap distance $d$ between two parallel electrode plates.
For many gases, the smallest breakdown voltages lie in the range of about \SIrange{10}{1000}{\Pa\cm}~\cite{izaMicroplasmasSourcesParticle2008}.
Reducing $d$ to \SI{100}{\um} or less allows plasma operation at or near atmospheric pressure ($p=\SI{101}{\kPa}$).\newline
The plasma setup presented in this work is a direct current (DC) microplasma where one of the electrodes is a nozzle with a small orifice through which gas is supplied (see Figure~\ref{fig:Plasma-Setup-Scheme}a in the Experimental section).
Whereas the geometry closely resembles that of a jet, the setup isn't technically defined as a plasma jet since the plasma is generated in the gap between the nozzle and the grounded electrode/sample~\cite{luAtmosphericpressureNonequilibriumPlasma2012}.
The interaction of plasmas with flat surfaces or nanoparticles is of interest for technical applications and a better understanding of plasma physics and chemistry.
Often, \emph{ex-situ} structural and chemical investigations on the milli- to nanometer scale are performed after plasma treatment of a material.
\newline
For these length scales, scanning electron microscopy is a valuable technique for microstructural and chemical investigations (typically using energy-dispersive x-ray spectroscopy, EDS).
Recently, the first microplasmas were generated inside scanning electron microscopes (SEMs)~\cite{matraLocalSputterEtching2013, muldersInsituLowEnergy2016a, pardinasDesignImplementationInSitu2016}.
A plasma-in-SEM setup not only reduces the time between plasma treatment and subsequent SEM analyses compared to a separate plasma setup, but also prevents exposure of the sample surface to ambient air.
The latter aspect enables studies of plasma-treated surfaces where subsequent contact with oxygen, humidity, or contamination must be avoided.\newline
Different approaches to generate plasmas in SEMs were demonstrated in earlier studies.
For example, local sputter etching was achieved by \citet{muldersInsituLowEnergy2016a} by introducing a small gas nozzle into an SEM and using the electron beam for ionization.
In the setup by these authors, the electron beam is scanned in a small slit in the nozzle near the orifice to generate ions in the gas stream.
The generated ions flow out of the orifice with the gas flow and are then accelerated toward the sample using an applied voltage between the nozzle and the stage. 
Modern SEMs often have a built-in option to apply the required negative voltage to the sample stage, typically used for beam-deceleration SEM imaging~\cite{phiferImprovingSEMImaging2009, jiruseNovelFieldEmission2014}. 
This approach does not require reaching the breakdown voltage of the gas, hence leading to a low-energy ion bombardment of the sample.
With this setup, low-energy \ce{Ar+} ions with energies ranging from \SIrange{20}{500}{eV} were used to remove amorphous surface layers~\cite{isaacsInsituLowEnergy2018, dutkaInsituLowEnergy2019}.\newline
Another plasma setup consists of a micro hollow cathode (or anode) DC plasma configuration in an environmental SEM (ESEM)~\cite{pardinasDesignImplementationInSitu2016}.
In the latter, the chamber pressure and gas type (in this case \ce{Ar}) is directly controlled with the ESEM.
A supplied high voltage (in this case \SI{+-1}{\kV}) generates the plasma, and the plasma-surface interaction subsequently can be analyzed within the ESEM.
Depending on the electrode polarity, either (i)~redeposition of sputtered material from the counter electrode onto the sample surface or (ii)~direct sputtering of the sample surface with positive ions was observed.
The sputtered area had a relatively large width of about \SI{2}{\mm}~\cite{pardinasDesignImplementationInSitu2016}.
A benefit of this experimental setup is that the gas-flow controls of the ESEM are used, which reduces the requirements for the hardware modifications to an SEM. 
However, a drawback is that using the low-vacuum mode reduces the image quality due to electron-beam scattering in the gas, resulting in a so-called electron-beam skirt \cite{danilatosFoundationsEnvironmentalScanning1988, goldsteinScanningElectronMicroscopy2018}.
This aspect impedes \emph{in-situ} SEM imaging of the plasma-sample interactions, limiting high-quality imaging to the normal high-vacuum mode of the ESEM.
To optimize image quality in gaseous environments, the distance between the end of the microscope’s pole piece and the sample, i.e., the gas-path length, is typically minimized to reduce the beam skirt. 
However, the gas-path length cannot be reduced too much for plasma experiments due to the risk of unwanted arcing to the microscope hardware.
Indeed, arcing from the micro hollow cathode to the microscope hardware over a relatively large distance of about \SI{25}{\mm} was reported for this setup using the low-vacuum mode~\cite{pardinasDesignImplementationInSitu2016}.\newline
\citeauthor{matraLocalSputterEtching2013}~\cite{matraCurrentVoltageCharacteristicsDC2013, matraLocalSputterEtching2013, matraCharacteristicsMicroPlasma2013} demonstrated a working jet-like microplasma setup inside an SEM.
This approach combines the properties of a jet (enabling a comparably high pressure in the gas jet compared to its environment) with the small dimensions of a microplasma for local plasma application (typically within a few ten \unit{\um}).
The gas flows from a gas nozzle with a small orifice (nominal diameter of a few ten \unit{\um}) toward a (flat) sample surface, whereas the chamber is continuously pumped to maintain a low overall pressure.
A plasma is generated by applying a voltage, here denoted as source voltage $V_\text{S}$, between the nozzle and the sample, somewhat similar to a plasma reactor with two electrode plates~\cite{gudmundssonFoundationsDCPlasma2017}.
However, the non-uniform pressure profile between the nozzle and the sample makes this plasma configuration unique, complicating the characterization of the plasma discharge.
The gap distance can be adjusted by using SEM imaging for alignment. 
The pressure profile between the nozzle and the sample can be modified by changing the gas flow, though it will also be heavily affected by the distance between the orifice and the sample.
A plasma is generated by applying at least the breakdown voltage between the nozzle and the sample (although the electron beam can be used to aid plasma ignition).
Depending on the gas, material removal by \ce{Ar+} sputtering~\cite{matraLocalSputterEtching2013} and growth of an \ce{C}-rich thin film~\cite{matraDCMicroplasmaJet2017} on a \ce{Si} surface were observed.
This proof-of-principle study~\cite{matraLocalSputterEtching2013} showed that a microplasma jet can be generated in the evacuated SEM chamber.\newline
However, the (desired) DC glow discharge was reported not to be fully stable, resulting in arcing to the sample~\cite{matraLocalSputterEtching2013} and a self-pulsing plasma mode for discharge currents in the range of about \SIrange{3}{30}{\uA} (depending on voltage, gas flow rate, and gap distance)~\cite{matraCurrentVoltageCharacteristicsDC2013}.
This arcing led to strong local heating and pronounced damaged spots on the sample~\cite{matraLocalSputterEtching2013}.
Furthermore, these previous studies did not investigate the possibility of \enquote{true} \emph{in-situ} SEM imaging, i.e., live SEM imaging during plasma operation. 
Instead, SEM images were taken before and after the plasma-treatment steps (also in ref.~\cite{pardinasDesignImplementationInSitu2016}), which will be denoted as \enquote{quasi} \emph{in-situ} operation in this work.
Still, these studies prove that a microplasma can be generated in an SEM and used for surface treatment.
This provides the opportunity to observe \emph{in-situ} changes of a sample's morphology and chemistry  on the \unit{mm} to \unit{nm} scale during plasma treatment using an SEM, ultimately leading to a better understanding of plasma-surface interactions and fundamental plasma properties.\newline
However, the availability of more studies is hampered by (i)~the required non-trivial modifications of an SEM and (2)~the lack of commercial solutions. 
In this work, a microplasma setup built inside a modern ESEM based on the work of \citet{matraLocalSputterEtching2013} is presented.
A stable operation of a DC discharge without arcing is realized.
Further, we present real-time \emph{in-situ} SEM imaging during plasma operation and show exemplary applications of our plasma-in-SEM setup for sputtering and local surface oxidation.
Finally, experimental challenges and potential upgrades of the setup are discussed.

\section{Results and discussion} \label{results}

The first part of this section shows results related to the microplasma and \emph{in-situ} SEM imaging. The second part discusses some exemplary results when applying the microplasma to materials. Finally, the third part reviews the limitations of this setup and proposes potential solutions to overcome these limitations.

\subsection{Microplasma Characterization}

\subsubsection{Gas-Pressure Profile}
\begin{figure*}[tb]
        
    \includegraphics[width=\linewidth]{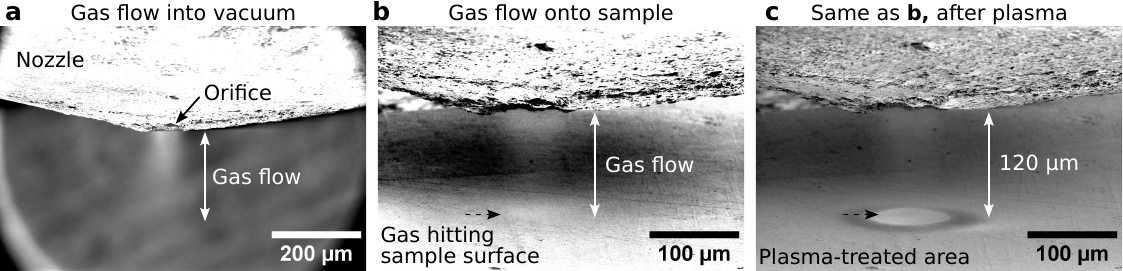}
    \caption{Investigation of the gas density profile in the plasma gap. \textbf{a}~SE-SEM image of the gas flow into vacuum acquired with a primary electron energy of \SI{2}{\keV}. \textbf{b}~A spot with a slightly increased SE signal is visible on the sample surface (marked with a dashed arrow) when a sample is brought into proximity, probably due to an increased gas density when the gas jet hits the sample surface. \textbf{c}~After plasma treatment, the bright spot coincides with the plasma-treated region, indicating that the gas spot in \textbf{b} can be used for aiming the microplasma at the desired region of interest.}
    \label{fig:Gas-Flow}
        
\end{figure*}
The used plasma setup has a non-uniform gas pressure along the plasma gap.
The gas density profile can be visualized by SEM imaging (Figure~\ref{fig:Gas-Flow}a) by using a low primary electron energy (here \SI{2}{keV}) to increase the electron-scattering probability and secondary electron (SE) generation within the gas cloud~\cite{muntzElectronBeamFluorescence1965}.
As a result, the SE-SEM image presumably shows higher intensity in regions with higher gas densities (Figure~\ref{fig:Gas-Flow}a).
Here, the gas cloud in Figure~\ref{fig:Gas-Flow}a flows into the microscope vacuum without obstruction. 
The contrast variations in the background result from out-of-focus imaging of the sample stage a few \unit{mm} below the nozzle along the electron-beam direction.
The gas density is highest close to the orifice and gradually decreases away from it.
This monotonic decrease is in accordance with calculated gas density profiles of restricted gas flows, e.g., in references~\cite{sharipovNumericalSimulationRarefied2004, misdanitisPressureDrivenRarefied2012, bykovBinaryGasMixture2020}. 
More explicitly, \citet{salehiCharacterization100Micrometerscale2019} report an exponential decay of the gas density away from an orifice from a simulation of gas jets for different pressure differences between the inside of the nozzle and the chamber. 
Experimental measurements of the pressure gradient away from the nozzle by \citet{patelFlowCharacterizationSupersonic2021} reveal a continuous pressure decrease away from the nozzle for a distance of about \num{20} orifice diameters (in their experiment about \SI{20}{\mm} for a \SI{0.8}{\mm} orifice diameter), which would correspond to a continuous pressure decrease away from the orifice of about \SI{400}{\um} for a nominal \SI{20}{\um} orifice diameter. 
From comparison with these results, we suspect a monotonic decrease in gas density and pressure across the microplasma gap in our experimental setup.\newline
However, if the gap distance is reduced by bringing the sample close to the orifice (here about \SI{120}{\um}), an increase in SE signal is visible on the sample surface as well (Figure~\ref{fig:Gas-Flow}b, dashed arrow).
The increased SE signal at the sample indicates an increased gas density at the sample surface.
From these observations, it becomes clear that the gas density profile in the gap depends, among other parameters, also on the gap distance.
This non-uniform gas pressure impedes predictions and comparison with conventional plasma reactors with a constant pressure between the electrodes.
As a beneficial side aspect, the visible gas spot on the sample surface can be used to predict the plasma-spot region.
This can be seen by comparing the images before and after plasma operation in Figures~\ref{fig:Gas-Flow}b and c, respectively, where the pit due to plasma sputtering forms in the region predicted in Figure~\ref{fig:Gas-Flow}b.

\subsubsection{Voltage-Current Characteristics of the Plasma}
\begin{figure*}[tb]
    \centering
        
    \includegraphics[width=\linewidth]{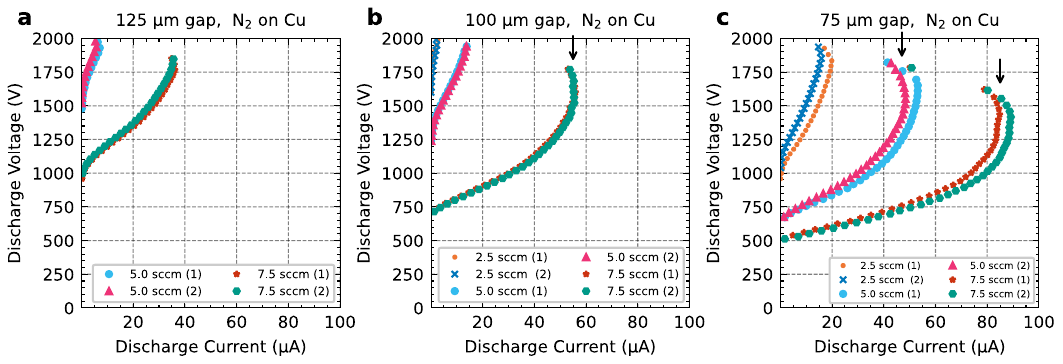}
    \caption{Voltage-current characteristics of a \ce{N2} microplasma for different gas flow rates and gap distances. All axis limits are equal for easier comparison. The data is shown for decreasing gap distance from left to right. Two measurements were performed for each gas flow rate and gap distance. No discharge was observed for \SI{125}{\um}/\SI{2.5}{\sccm}. Slight deviations between these measurements are mainly caused by uncertainties in gap distance. The apparent drop in current for higher voltages (marked with arrows in b and c) is a measurement artifact caused by sample-surface sputtering. In general, larger discharge currents are observed for higher gas flow rates and smaller gap distances. A positive slope for all curves indicates a so-called abnormal glow discharge behavior.}
    \label{fig:VI-Curves-N2-on-Cu}
        
\end{figure*}
Next, the voltage-current characteristics (i.e., the dependence of discharge voltage $V_\text{D}$ and discharge current $I_\text{D}$) of a \ce{N2} microplasma were investigated for three gap distances (\SI{75}{\um}, \SI{100}{\um}, and \SI{125}{\um}) and three gas flow rates (\SI{2.5}{\sccm}, \SI{5.0}{\sccm}, and \SI{7.5}{\sccm}).
Nitrogen was chosen over \ce{Ar} because it resulted in lower chamber pressures for the same gas flow rate, allowing for higher gas flow rates (up to \SI{8}{\sccm}) into the high-vacuum microscope chamber. 
The pumping speed of different gases is discussed in more detail in the supplementary information.
For each gas-flow/gap-distance pair, two measurements were taken for repeatability (here denoted in the brackets in the figure legends).
Figures~\ref{fig:VI-Curves-N2-on-Cu}a--c show the values sorted with decreasing gap distance from left to right.
The same axis limits were used for easier comparison.

In general, a positive slope is visible for all curves, indicative of a so-called abnormal glow discharge plasma~\cite{gudmundssonFoundationsDCPlasma2017}.
This was also observed by \citet{matraCurrentVoltageCharacteristicsDC2013}, but not in all of their measurements.
After this initial positive increase of discharge current with discharge voltage, nearly all curves show a maximum current followed by a  current decrease (cf. arrows in Figures~\ref{fig:VI-Curves-N2-on-Cu}b and c).
The last aspect is a measurement artifact, probably caused by rapid sputtering of the electrode, and should not be interpreted as an actual voltage-current characteristic of the microplasma.
This artifact is discussed in more detail in the supplementary information (Figure~\ref{fig:suppl:SI-Problems-VI-Measurements}).\newline
Next, the ordinate intercepts of the curves in Figure~\ref{fig:VI-Curves-N2-on-Cu} are discussed. 
These points correspond to the lowest discharge voltage at which a plasma discharge can be sustained. 
Note that this isn't equal to the breakdown voltage, as the voltage required to initiate a breakdown is often (significantly) higher than the voltage required to sustain one~\cite{fridmanPlasmaPhysicsEngineering2011}.
The actual breakdown voltages were not measured since our setup does not produce the necessary uniform gas pressure for a given gap distance for a controlled measurement~\cite{lisovskiyLowpressureGasBreakdown2000}. 
Figures~\ref{fig:VI-Curves-N2-on-Cu}a--c show a decreasing minimum discharge voltage for increasing gas flow rates for the same gap distance. 
Since an increase in gas flow rate for a constant gap distance is assumed to result in an increasing gas density, this decreasing minimum discharge voltage offers an interesting insight into the plasma discharge. 
As described by the Paschen curve for simple parallel-plate and uniform-pressure DC plasma systems, an increased pressure heavily affects the discharge properties (see also supplementary Figure~\ref{fig:suppl:SI-PaschenCurve-Nitrogen}).\newline
On the one hand, if the gas density is higher than the optimum (i.e., the point with the lowest minimum discharge voltage, similar to the minimum in the Paschen curve), the electrons undergo many collisions, which limit their possibility to gain enough energy to ionize a molecule. 
This ionization is required to create an avalanche effect, which is needed to sustain a discharge. 
In this case, a higher voltage is required to sustain the discharge to ensure the electrons can gain sufficient energy to cause subsequent ionization.\newline
On the other hand, if the gas density is lower than the optimum, the electrons can easily gain sufficient energy, but they may not collide frequently enough to cause the further ionization required to sustain the discharge. 
Then, again, a higher voltage is required to ensure that the collisions will cause ionization.
As the minimum voltage required to sustain a discharge decreases with increasing gas density, it is implied that the gas density is lower than the optimal case overall. 
This is analogous to being on the left side of the minimum in the Paschen curve. 
It should be noted, though, that given the strong pressure gradient in this setup, the discharge mechanisms are not as straightforward as they are assumed by the Paschen curve, so a direct comparison is difficult. 
This behavior of the minimum discharge voltage implies that the plasma could be categorized as a so-called obstructed abnormal glow discharge~\cite{gudmundssonFoundationsDCPlasma2017, fridmanPlasmaPhysicsEngineering2011}.

When comparing the curves for the same gas flow rate and different gap distances in Figures~\ref{fig:VI-Curves-N2-on-Cu}a-c, both the gas density and the gap distance are varied since the former is affected by the latter.
Assuming that the gas density at a constant gas flow rate increases for a decreasing gap distance, the changes in minimum discharge voltage in Figures~\ref{fig:VI-Curves-N2-on-Cu}a-c indicate that the gas density is increasing non-linearly (in contrast to linearly decreasing distance) and more substantial than the gap distance. 

An additional complication affecting the interpretation of the data is the setup geometry.
The shown setup with a rounded nozzle with an orifice as one electrode and a possibly textured sample surface as another electrode is different from earlier publications studying various electrode geometries~\cite{lisovskiyLowpressureGasBreakdown2000, lisovskiyElectricFieldNonuniformity2017, mathewExperimentalVerificationModified2019, fuElectricalBreakdownMacro2020}.
Microplasmas are especially sensitive to surface effects due to the small spatial scale in the sub-\unit{mm} range, as the electric field can be strongly altered by small morphological changes in the electrode surfaces~\cite{fuElectricalBreakdownMacro2020}.
In addition, due to the high-pressure gradient, it is impossible to accurately control the pressure in the discharge gap using this setup. 

\subsubsection{Plasma Generation and Stability in an SEM}
The plasma-in-SEM setup enables studying the interplay between the electron beam of the SEM and the plasma.
Different aspects of this interaction are discussed in the following.

\begin{figure*}[tb]
        
    \centering
    \includegraphics[width=\linewidth]{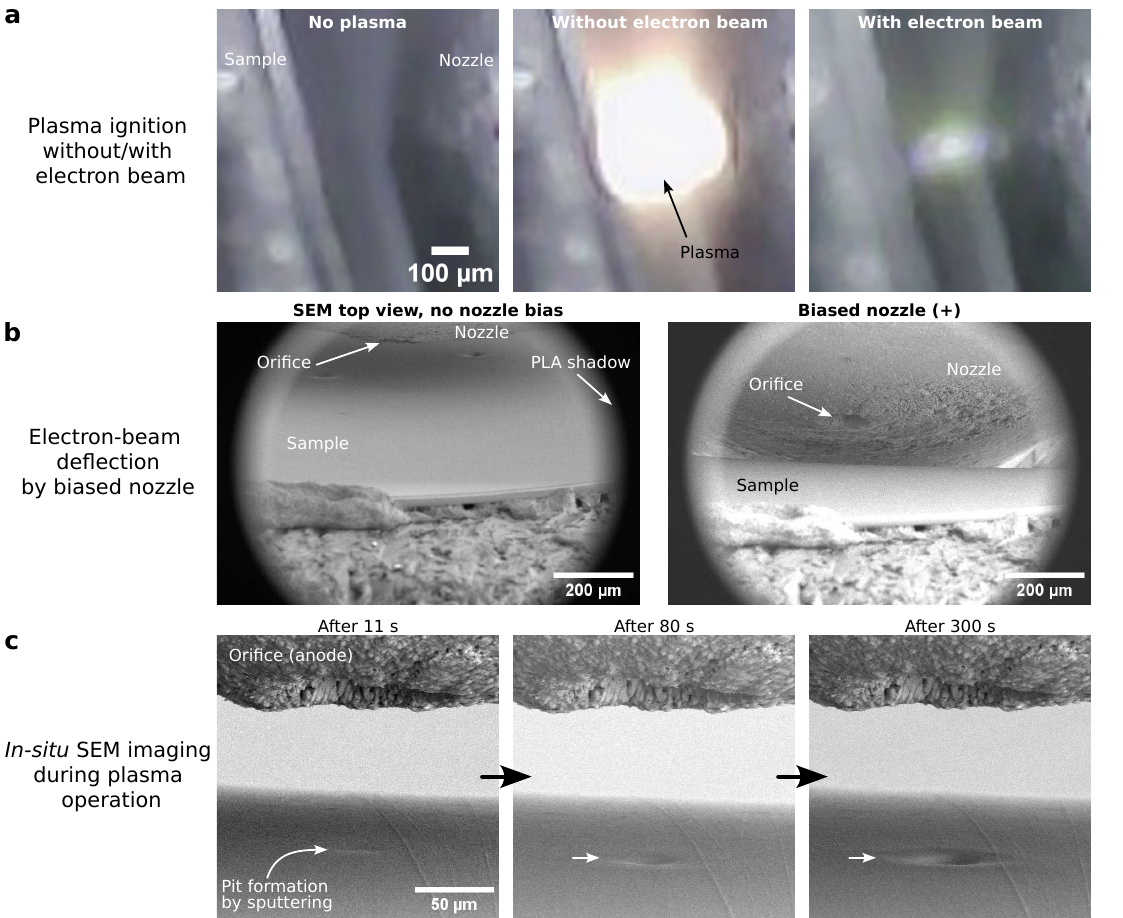}
    \caption{Aspects of microplasma operation in a scanning electron microscope. \textbf{a}~Webcam images of plasma operation. The electron beam can be used to ignite the plasma at a lower applied source voltage to generate a less intense plasma (right) compared to self-ignition by reaching the breakdown voltage (middle). The shown plasma images correspond to the plasma conditions right after plasma ignition. \textbf{b}~Top-view SE-SEM images (\SI{10}{\keV}) of the nozzle and sample (left) without and (right) with applied voltage on the nozzle ($V_\text{S} = \SI{920}{\V}$). In this example, the electrons are attracted to the positive potential on the nozzle, which enables imaging of the orifice area. \textbf{c}~True \emph{in-situ} SE-SEM imaging during plasma operation is possible and shows the formation of a pit in the sample due to sputtering.}
    \label{fig:Plasma-in-SEM-Examples}
        
\end{figure*}

Firstly, an electron beam can be used to ignite the plasma at lower voltages than required for the self-ignition when reaching the breakdown voltage~\cite{matraLocalSputterEtching2013} (Figure~\ref{fig:Plasma-in-SEM-Examples}a). 
For example, in one case a plasma discharge could not be achieved, even when applying a maximum source voltage $V_\text{S} = \SI{2}{\kV}$ to the nozzle without an electron beam. However, scanning with the electron beam caused a plasma discharge already at $V_\text{S} = \SI{920}{\V}$ for the same gap distance and gas flow rate.
This can be explained by the generation of SEs, backscattered electrons (BSEs), and x-rays upon the interaction of the electron beam with the sample, which then triggers the plasma ignition.
Notably, the electron beam ignites the plasma even if not directly scanning in the gap region.
A webcam video comparing plasma ignition by (i)~reaching breakdown voltage (the conventional way) or (ii)~using the electron beam is found in the supplementary information (\emph{Breakdown-vs-SEM-Plasma.mp4}).
In this video, the SEM-triggered plasma shows a less intense plasma cloud than the self-ignited plasma.
Therefore, the electron beam can be advantageously employed to ignite a less intense plasma at lower voltages (cf. middle and right images in Figure~\ref{fig:Plasma-in-SEM-Examples}a).
In addition, for conditions where a plasma is not self-sustainable, i.e., with a large gap distance and/or low gas flow rate, a plasma discharge was observed that was only active during active electron-beam scanning (see supplementary information Figure~\ref{fig:suppl:SI-BeamCurrentPlasma}).

Secondly, applying an electric potential to the nozzle will create an electric field that deflects the incoming electron beam, e.g., toward the positive potential on the nozzle (Figure~\ref{fig:Plasma-in-SEM-Examples}b).
The deflection depends on the electron energy (less deflection for higher \unit{\keV}) and probably also on the extent of the exposed metal part of the steel nozzle. 
In our setup, insulating tape was used to cover most of the steel nozzle, excluding the tip (see black tape in Figure~\ref{fig:Plasma-Setup-Scheme}). 
The deflection may be minimized by (i)~shielding the open metallic surface of the nozzle tip and (ii)~using a higher primary electron energy.
However, the deflection can also be used advantageously. 
For example, the deflection can be strong enough so that the SE-SEM image is formed from the nozzle-tip surface, e.g., at $V_\text{S} = \SI{920}{\V}$ for a primary beam energy of \SI{10}{\keV} (Figure~\ref{fig:Plasma-in-SEM-Examples}b, right).
In this way, the tip region of the nozzle can be imaged with the SEM even though it is aligned parallel to the electron beam, i.e., without a direct line of sight.
This effect is more pronounced at lower electron energies.
A supplementary movie (\emph{SEM\_Plasma\_Ignition.mp4}) shows correlative imaging of the webcam and SEM images during a gradual increase in the source voltage $V_\text{S}$ and subsequent SEM-induced plasma ignition. 
The SEM image is increasingly \enquote{tilted} toward the nozzle with increasing $V_\text{S}$.

Thirdly, it was observed that \emph{in-situ} SEM imaging during plasma operation is indeed possible, opening up the opportunity for time-resolved studies.
In SE-SEM imaging, a working plasma leads to an increase in signal (brightness) using the Everhart-Thornley detector (ETD). 
For imaging, this effect can be compensated by reducing the ETD bias setting. 
For a \ce{CO2} plasma, this method proved effective for discharge currents up to about \SI{7}{\uA}, after which the ETD was saturated (i.e., no further reduction in bias possible), and no SE-SEM imaging was possible.
It is remarkable that \emph{in-situ} SE-SEM imaging during plasma operation is feasible, despite several challenges: (i)~the electron-beam current used (few \unit{\nA}) is about a thousand times lower than the measured discharge current (few \unit{\uA}), (ii)~many spurious SEs are likely generated in the plasma region~\cite{gudmundssonFoundationsDCPlasma2017, fridmanPlasmaPhysicsEngineering2011}, and (iii)~the positive suction voltage on the ETD of \SI{+250}{\V} to attract SEs is comparatively low compared to the nozzle voltage (typically \SI{>1}{\kV}).
For example, three SE-SEM images taken during continuous microplasma operation are shown in Figure~\ref{fig:Plasma-in-SEM-Examples}c.
The plasma duration increases from left to right, leading to increasing pit diameter and depth due to surface sputtering.
The most notable distortion in the SE-SEM image is caused by the applied nozzle voltage, resulting in an electron-beam deflection (Figure~\ref{fig:Plasma-in-SEM-Examples}b).
\newline
Similarly, BSE-SEM imaging was tested by negatively biasing the ETD with \SI{-150}{\V} to suppress (mainly) SEs from the image signal. 
In contrast to SE-SEM imaging, the BSE-SEM image brightness is not affected by the discharge current during plasma operation, meaning that BSE-SEM imaging is still possible even when the SE signal becomes saturated at high discharge currents (e.g., \SI{>7}{\uA} for \ce{CO2}).
A video comparing BSE- and SE-SEM imaging is found in the supplementary information (\emph{In-Situ-SEM\_SE-vs-BSE.mp4}).
Since the ETD covers only a relatively small solid angle, it is inefficient for BSE detection.
This results in a lower signal yield than for SE-SEM imaging.
However, the low BSE signal may be increased by using a more efficient and low-vacuum compatible BSE detector~\cite{stokesEnvironmentalScanningElectron2012, nedelaHighefficiencyDetectorSecondary2018}, but this was not tested in this work.
Since both SE- and BSE-SEM imaging is possible and similar to conventional SEM imaging, the signals can be chosen depending on the experiment, or both signals can be collected with two different detectors.
This enables more surface-sensitive imaging with SEs and $Z$-dependent imaging with BSEs~\cite{goldsteinScanningElectronMicroscopy2018}.

An application-relevant observation from the demonstrated setup is the absence of undesired high-current and high-frequency arc discharges, which were reported by \citet{matraCurrentVoltageCharacteristicsDC2013} as a self-pulsating plasma mode.
Instead, we observed stable DC glow discharges with discharge currents ranging from about \SIrange{0.1}{175}{\uA}, which can be controlled by adjusting $V_\text{S}$. 
This corresponds to current densities ranging from \SI{5}{\mA.\cm^{-2}} to \SI{9}{\A.\cm^{-2}} for an assumed plasma-spot diameter of \SI{50}{\um}.
The latter can vary depending on the gap distance.
We did not investigate higher currents than \SI{175}{\uA} since the \SI{30}{\um} thick \ce{Cu} target is sputtered away in a few (ten) seconds at the plasma spot. 
Conversely, the plasma could not be sustained below the lower limit of about \SI{0.1}{\uA}.\newline
The absence of arcing may be explained by the lower chamber pressure in our used SEM (about \SI{2e-2}{\Pa}) compared to the reported values \enquote{below \SI{1}{\Pa}}~\cite{matraCurrentVoltageCharacteristicsDC2013}. 
Notably, a self-pulsing plasma was observed for the shown setup when powering it in ambient air during prototyping.
The high-frequency arcing in this self-pulsing mode (a few ten \unit{\kilo\hertz}) causes significant electromagnetic interference to surrounding electronic devices, including the SEM.
In addition, powering the setup in the low-vacuum mode of the SEM at a chamber pressure of \SI{40}{\Pa} leads to undesired discharges in the SEM chamber, similarly as observed by \citet{pardinasDesignImplementationInSitu2016}.
This restricts the plasma operation to the high-vacuum mode (below \SI{3.3e-2}{\Pa} for the used SEM).
Here, only occasional arcs during plasma operation were observed when non-flat samples with surface protrusions were used.
It may be possible to fully mitigate the self-pulsing plasma mode by an optimal choice of electronic components in the circuit. 
Still, in our case, the reduced chamber pressure (about \SI{2e-2}{\Pa}) is the most likely reason for a stable DC plasma operation compared to \citet{matraCurrentVoltageCharacteristicsDC2013}.

\subsection{Microplasma Applications}

\subsubsection{Sputtering and Cone Formation}

Sputtering is the process of removing atoms of the target material by impinging ions.
Sample material was removed by this process in all experiments, where the sample was used as the cathode. 
The positively charged ions are accelerated toward the cathode and cause sputtering, as is common in glow discharges.
This results in changes in surface morphology in the plasma-spot regions, with diameters ranging from about \SIrange{50}{150}{\um} (depending on the gap distance, pressure, discharge voltage/current, and plasma duration).
In the following, results for sputtering on (i)~a polished or (ii)~a Ni nanoparticle-covered \ce{Cu} surface are shown.
\begin{figure*}[tb]
        
    \includegraphics[width=\linewidth]{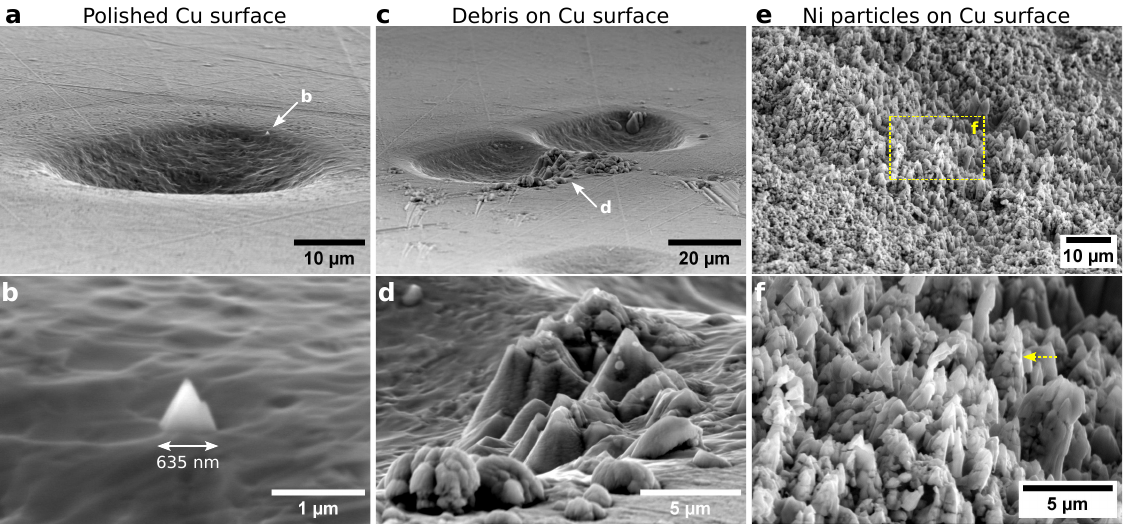}
    \caption{Cone formation after \ce{Ar^+}-ion sputtering for different concentrations of surface particles. The lower row shows higher magnification SEM images of the upper row. \textbf{a}~Cone formation is not visible in the shown region of a polished \ce{Cu} surface. A small cone is visible on the edge (\textbf{b}), probably due to a small contaminating particle on the sample surface. \textbf{c} and \textbf{d}~Debris on the \ce{Cu} surface forms cones under plasma treatment. \textbf{e}~\ce{Ni} particles deposited on a \ce{Cu} substrate show clear cone formation in the plasma-treated region. In the early stages of sputtering, the \ce{Ni} particles locally agglomerate to form a cone (see the example in \textbf{f} marked with a dashed arrow).}
    \label{fig:Cone-formation}
        
\end{figure*}

The formation of a pit under \ce{CO2} and \ce{Ar}-containing plasma was observed for a polished \ce{Cu} surface.
An example is shown in Figure~\ref{fig:Cone-formation}a, which was created with \ce{Ar} plasma.
Experimentally, this pit formed after \SI{1.2}{\keV} \ce{Ar^+} exposure with a discharge current of about \SI{15}{\uA} (current density of \SI{1.2}{\A.\cm^{-2}} for a plasma-spot pit diameter of \SI{40}{\um})  for about \SI{10}{\s}. 
The rapid pit formation is indicative of the high sputter rates of the setup. 
The pit surfaces are rougher than the original polished surface. 
A comparably small conical structure is visible at the edge of the pit, which is magnified in Figure~\ref{fig:Cone-formation}b.
This may have been an impurity or other contamination present in or on the \ce{Cu} surface, which deformed to the shown conical structure during sputtering.
Its bright appearance in the SE-SEM images may be explained by the penetration depth of primary electrons, here at an energy of \SI{15}{\keV}. 
For relatively thin structures such as the shown impurity in Figure~\ref{fig:Cone-formation}b, SEs are emitted not only on the entrance surface of the beam but also on the exit surface of the cone (and also the sample material behind the cone). 
The additional SE emission from the exit surface (relative to the incoming electron-beam direction) leads to higher SE-SEM image intensity for thinner sample regions.

\begin{figure*}[tb]
        
    \includegraphics[width=\linewidth]{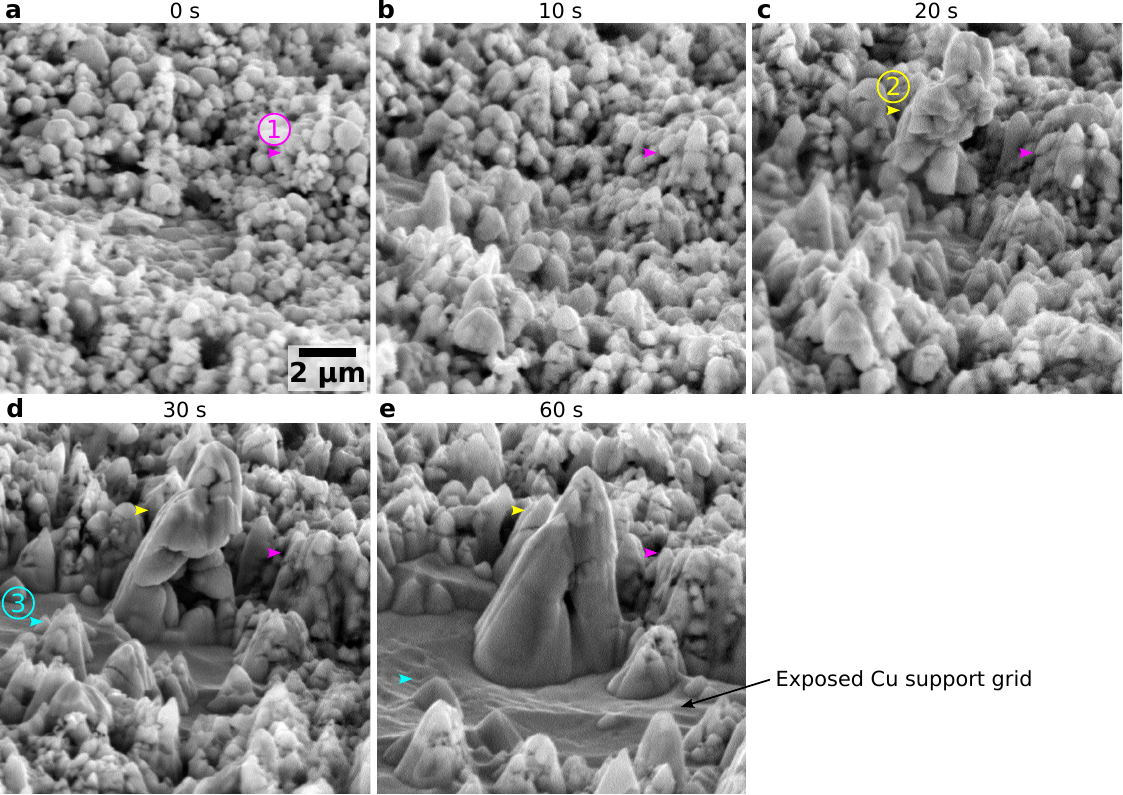}
    \caption{Quasi \emph{in-situ} observation of \ce{Ni} nanoparticle agglomeration and subsequent cone formation during \ce{Ar^+}-ion sputtering (\SI{5}{\uA}, \SI{1.32}{keV}) for the given duration shown above the SE-SEM images. The latter were taken with a large-field detector (LFD) in the low-vacuum mode (\SI{40}{\Pa}) after each plasma operation in high-vacuum mode. A few interesting regions are marked with arrows. Region (1)~shows gradual change from (\textbf{a}) a round particle morphology to an increasing distortion toward a conical shape. After reaching the latter in \textbf{d}, the cone is then starting to be removed by sputtering, as visible in \textbf{e}. Region (2) shows the sudden agglomeration of a few nanoparticles in \textbf{c}. A larger cone is forming from this agglomeration (\textbf{d} and \textbf{e}). Region (3) exemplifies that, after initial formation, the cones are sputtered away under further \ce{Ar^+}-ion bombardment (\textbf{d} and \textbf{e}).}
    \label{fig:Cone-formation-2}
        
\end{figure*}

Cone formation is observed for random debris (Figures~\ref{fig:Cone-formation}c and d) or full coverage with \ce{Ni} nanoparticles (Figures~\ref{fig:Cone-formation}e and f).
For the latter, the nanoparticles seem first to cluster together (see region marked with an arrow in Figure~\ref{fig:Cone-formation}f) and then tend toward a conical shape during prolonged sputtering.
The latter aspect was studied in more detail by monitoring the same area after a certain plasma duration with SEM imaging (Figure~\ref{fig:Cone-formation-2}). 
Between each plasma treatment, the sample area was moved onto the optical axis of the SEM to allow for high-magnification imaging.
A few exemplary areas are annotated with arrowheads, which are discussed in the following.\newline
Region (1) shows the shape evolution of \ce{Ni} particles under \ce{Ar} plasma. 
Between \SIrange{0}{30}{\s}, the shape gradually changes from round nanoparticles (Figure~\ref{fig:Cone-formation-2}a) toward a conical shape (Figure~\ref{fig:Cone-formation-2}d). 
After reaching the final conical shape, the cone is milled away during further sputtering (cf. region (1) for Figures~\ref{fig:Cone-formation-2}d and e).
Similarly, region (3) shows the removal of smaller cones between \SIrange{30}{60}{\s} of sputtering (cf. region (3) for Figures~\ref{fig:Cone-formation-2}d and e).
For region (2), a few \ce{Ni} nanoparticles agglomerate between \SIrange{10}{20}{\s}.
It is unclear from the images if these particular particles result from the present particles in the shown region or were redeposited from remote sample areas.
Further plasma exposure leads to a merging of the individual nanoparticles and the formation of a larger cone with smooth surfaces (cf. region (2) for Figures~\ref{fig:Cone-formation-2}d and e).
Overall, after \SI{60}{\s}, the underlying \ce{Cu} surface is partly exposed (Figure~\ref{fig:Cone-formation-2}e).

The observed formation of cones is a commonly observed modification of metal surfaces under ion bombardment~\cite{guentherschulzeNeueUntersuchungenUeber1942, aucielloIonInteractionSolids1981, wehnerConeFormationResult1985}.
The cone shape is commonly thought of as a combined result of varying sputter yield depending on the (i)~ion-incidence angle and (ii)~material. 
The sputter yield typically increases with increasing ion-incidence angle up until a maximum value $\Theta_\text{m}$, and then decreases rapidly toward grazing incidence (i.e., the ion direction being parallel to the sample surface)~\cite{aucielloIonInteractionSolids1981}.
This results in a cone shape of impurities and surface particles before complete removal by sputtering.
The seeds for the cones can be intrinsic elemental impurities in an otherwise flat surface or particles on the surface with lower sputter yield. 
The latter correlates with the melting temperature of a material. \citet{wehnerConeFormationResult1985} has tested numerous surface/seed combinations of metals with different melting temperatures and found that cone formation requires seed materials with higher melting temperatures than the surface material.
This is the case for \ce{Ni} particles ($T_\text{melt} = \SI{1728}{\K}$) on a \ce{Cu} substrate ($T_\text{melt} = \SI{1358}{\K}$) observed in Figure~\ref{fig:Cone-formation-2}.
Note that the used nanoparticles are large enough (around \SI{100}{\nm}) so that a reduction in melting points is assumed to be negligible~\cite{teijlingenSizedependentMeltingPoint2020,pabariSizeDependentProperties2022}.\newline
A mean apex angle of $\Theta = \SI[separate-uncertainty = true]{61.4(11.1)}{\degree}$ (the error being the standard deviation) was measured for \num{40} cones.
According to \citet{stewartMicrotopographySurfacesEroded1969}, $\Theta$ is related to the ion-incidence angle for maximum sputter yield $\Theta_\text{m}$, as $\Theta_\text{m} = \left(\SI{180}{\degree} - \Theta\right)/2 = \SI[separate-uncertainty = true]{59.3(5.6)}{\degree}$.
The experimental value $\Theta$ is in good agreement with the maximum $\Theta_\text{m,sim} \approx \SI{65}{\degree}$ of a simulation of the angle dependence of the sputter yield of \ce{Ar} on \ce{Ni} using SRIM (supplementary information Figure~\ref{fig:suppl:SRIM-Y-angle-sim}). The differences between measured and simulated values can be explained by (i)~limited statistics based on only \num{40} measured cones, (ii)~systematic errors in the angle measurement from SEM images, and (iii)~uncertainties in the simulation~\cite{shulgaNoteArtefactsSRIM2018}.

\subsubsection{Local Oxidation}
\begin{figure*}[tb]
        
    \includegraphics[width=\linewidth]{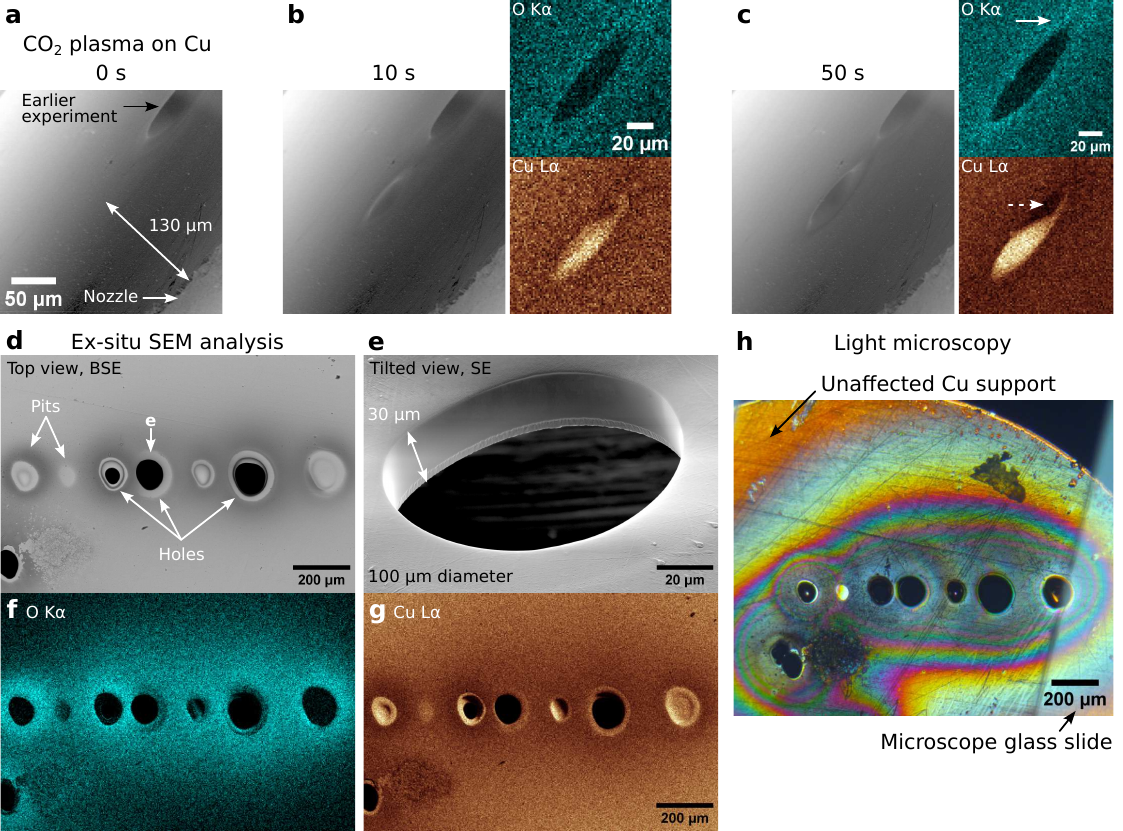}
    \caption{Sputtering and oxidation of a polished \ce{Cu} surface under \ce{CO2} plasma. \textbf{a}~SE-SEM image showing the sample surface opposite to the nozzle with a \SI{130}{\um} gap. The hole in the top-right corner is from an earlier experiment. \textbf{b and c}~Images and \ce{O}/\ce{Cu} elemental maps after \SI{10}{\s} and \SI{50}{\s} plasma treatment. A pit forms due to sputtering. A higher \ce{Cu} signal in the pit indicates the removal of the native oxide in the plasma spot. \textbf{c}~Enhanced \ce{O} signal is visible at the pit's edge (marked with a solid arrow). The depletion of \ce{Cu} signal is due to the shadowing of the x-ray signal toward the detector. \textbf{d}~Top-view BSE-SEM image of various pits and holes in the \ce{Cu} foil after plasma treatment. \textbf{e}~Side-view SE-SEM image of a hole showing vertical side walls. \textbf{f and g}~The elemental maps reveal enhanced oxidation around the plasma spots and higher \ce{Cu} signal in the pits similar to \textbf{b} and \textbf{c}. \textbf{h}~Light-microscopy image showing interference effects in the oxidized regions around the plasma spots.}
    \label{fig:Cu-oxidation-CO2}
        
\end{figure*}

Plasma finds applications in both the oxidation and reduction of materials~\cite{wangCatalystPreparationPlasmas2018, yeReviewAdvancesCatalyst2022}.
Here, we investigate the possibilities of local plasma-induced sample oxidation in the SEM.
As a first example, a polished \ce{Cu} surface was exposed to a \ce{CO2} plasma (Figure~\ref{fig:Cu-oxidation-CO2}).
The gap distance was approximately \SI{130}{\um} (Figure~\ref{fig:Cu-oxidation-CO2}a).
In Figure~\ref{fig:Cu-oxidation-CO2}a, a sputtered hole from a previous experiment is visible in the top right corner, and the nozzle is visible in the bottom right corner.
The applied source voltage was $V_\text{S} = \SI{2}{\kV}$ and discharge currents between \SIrange{70}{120}{\uA} were measured.
After \SI{10}{\s} of plasma operation, Figure~\ref{fig:Cu-oxidation-CO2}b, a pit starts forming with a diameter of about \SI{70}{\um}.
Chemical analysis by EDS shows increased \ce{Cu} and decreased \ce{O} signals in the pit region, indicating a removal of the native \ce{Cu} oxide by sputtering.
This exposes the underlying \ce{Cu} metal, leading to a higher \ce{Cu}~L$\alpha$ signal.
After \SI{50}{\s}, the pit is widened to about \SI{100}{\um} diameter (Figure~\ref{fig:Cu-oxidation-CO2}c).
The sputtered pit area still shows a higher \ce{Cu} signal than the unaffected \ce{Cu} surface around it, similar to Figure~\ref{fig:Cu-oxidation-CO2}b. 
The reduction in \ce{Cu}~L$\alpha$ signal in the top part of the \ce{Cu} elemental map in Figure~\ref{fig:Cu-oxidation-CO2}c results from shadowing of the generated \ce{Cu}~L$\alpha$ signal x-rays from the inside of the pit toward the EDS detector.
An increase in \ce{O}~K$\alpha$ signal is visible at the pit's edge (Figure~\ref{fig:Cu-oxidation-CO2}c).
This observation indicates the oxidation of \ce{Cu} in this region.
The \ce{O} signal increases under prolonged \ce{CO2}-plasma exposure (not shown here), which we attribute to the continuous growth of this \ce{Cu}-oxide layer.

The sample was investigated again in the SEM and using light microscopy after the \emph{in-situ} experiments (Figures~\ref{fig:Cu-oxidation-CO2}d--h).
The top-view BSE-SEM image acquired at \SI{20}{\keV} shows different experimental sites of local \ce{CO2} plasma treatment (Figure~\ref{fig:Cu-oxidation-CO2}d). 
The black areas show regions where the total thickness (about \SI{30}{\um}) of the \ce{Cu} support was sputtered away, leaving holes behind.
One of the holes is also displayed in the SE-SEM image in Figure~\ref{fig:Cu-oxidation-CO2}e.
The tilted view reveals the high aspect ratio of the sputtering process, resulting in vertical sidewalls.
The elemental map of \ce{O} shows an increased \ce{O} signal around the plasma spots, similar to Figure~\ref{fig:Cu-oxidation-CO2}c, which is decreasing in radial direction away from the spots.
For the pits, the removal of the native oxide layer of \ce{Cu} leads to an increased \ce{Cu}~L$\alpha$ signal.
The increased \ce{O} concentration around the holes reduces the effective atomic number relative to metallic \ce{Cu}.
This results in a reduced BSE intensity in Figure~\ref{fig:Cu-oxidation-CO2}d in the oxidized regions due to the BSE signal's atomic number $Z$ dependence~\cite{goldsteinScanningElectronMicroscopy2018}.
Interestingly, the oxidation of the \ce{Cu} surface reaches a few hundred \unit{\um} away from the initial plasma spots. 
This phenomenon is more clearly visible in the light-microscopy image (Figure~\ref{fig:Cu-oxidation-CO2}h), which shows interference effects related to the gradually changing thickness of the grown \ce{Cu}-oxide film (Newton rings).
In the top left corner of the image, there is an unaffected (i.e., without plasma-induced oxidation) area of the sample (marked with an arrow in Figure~\ref{fig:Cu-oxidation-CO2}h).
Overall, the polished \ce{Cu} surface is sputtered away under \ce{CO2} plasma. 
A local \ce{CO2} plasma causes oxidation around the plasma spot, probably forming a \ce{Cu}-oxide film with decreasing thickness away from the plasma spot. 
This oxidation is most likely caused by oxygen species (such as atomic or ionized \ce{O}) generated in the plasma. 
These species can be transported out of the plasma (so-called afterglow) by the gas flow, explaining why the oxidation of the \ce{Cu} is observed away from the plasma spot as well. 

\begin{figure*}[tb]
        
    \includegraphics[width=\linewidth]{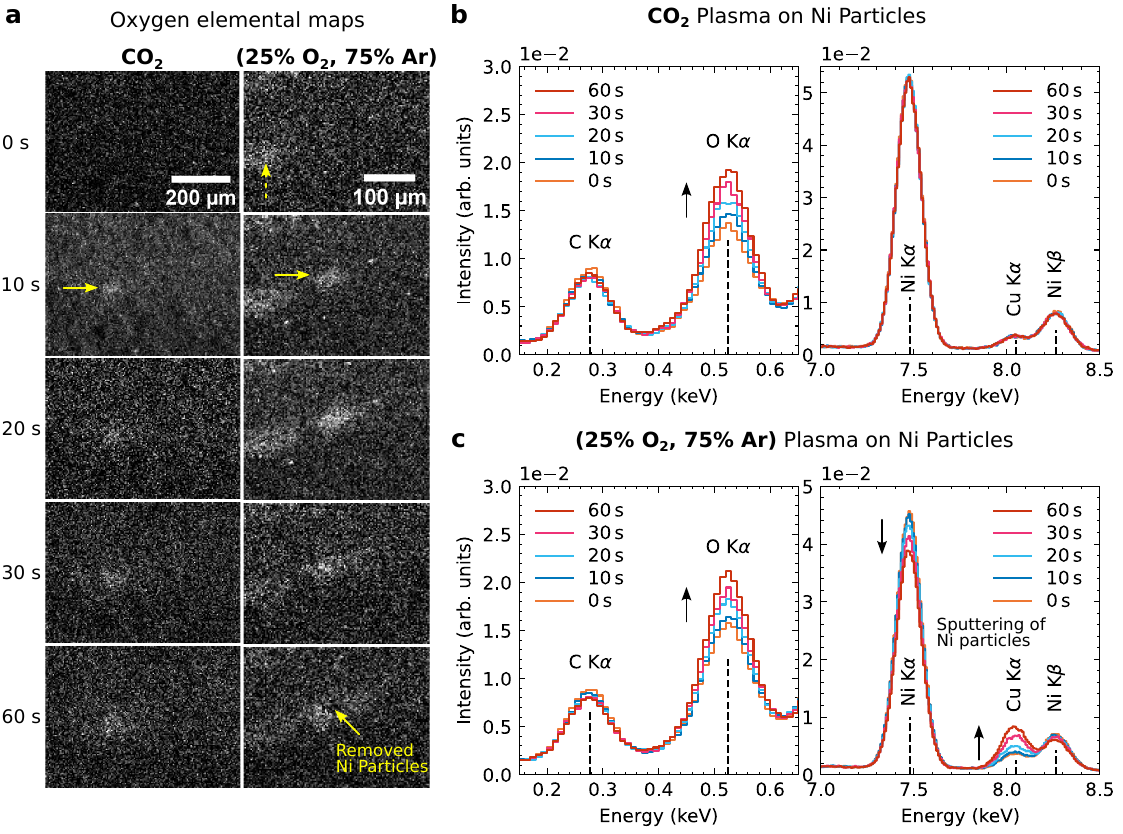}
    \caption{Local oxidation of \ce{Ni} particles under \ce{CO2} and \ce{O2}/\ce{Ar} plasma treatment. \textbf{a}~Elemental maps showing the \ce{O}~K$\alpha$ intensity for increasing plasma duration between \SIrange{0}{60}{\s} (top to bottom) for \ce{CO2} plasma (left column) and \ce{O2}/\ce{Ar} plasma (right column) for similar discharge current (about \SI{5}{\uA}) and gap distance (about \SI{250}{\um}). A spot of local oxidation is visible after \SI{10}{\s} (marked with horizontal arrows). The \ce{O}-rich spot at \SI{0}{\s} for \ce{O2}/\ce{Ar} is from a previous experiment (dashed vertical arrow). (\textbf{b}~and \textbf{c})~Comparison of extracted EDS signals in selected energy region for the \ce{O} (left), and \ce{Ni} and \ce{Cu} energy regions (right). The increase in \ce{O} signal for increasing plasma duration is visible. \textbf{c}~For \ce{O2}/\ce{Ar} plasma, sputtering of \ce{Ni} particles and subsequent exposition of the underlying \ce{Cu} support reduces the \ce{Ni}~K$\alpha$ signal and increases the \ce{Cu}~K$\alpha$ signal. This effect is absent in \textbf{b}, indicating a significantly reduced sputter yield for \ce{CO2} plasma. For comparison, the EDS spectra in \textbf{b} and \textbf{c} were normalized to the integrated signals in the energy intervals $[\SI{2}{\keV}, \SI{5}{\keV}]$ and $[\SI{10}{\keV}, \SI{14}{\keV}]$ containing only  bremsstrahlung background signal.}
    \label{fig:Ni-oxidation-comparison-gases}
        
\end{figure*}

Next, similar experiments with \ce{CO2} plasma on \ce{Ni} nanoparticles were performed (Figure~\ref{fig:Ni-oxidation-comparison-gases}a, left column).
The \ce{Ni} particles were deposited on a \ce{Cu} support film and formed a layer with a (varying) thickness of a few \unit{\um} (Figure~\ref{fig:Cone-formation}e).
The gap distance was \SI{250}{\um}, and the discharge current was \SI{5}{\uA}.
Local oxidation was observed \emph{inside} the plasma spot, as marked by the arrow in the elemental map acquired after \SI{10}{\s} plasma exposure.
The \ce{O} signal increases with increasing plasma duration from \SIrange{0}{60}{\s}. 
This aspect is not as evident in the noisy elemental maps but more clearly visible in the summed-up and normalized EDS spectra (see Figure~\ref{fig:suppl:EDS-Normalization} for details) from the plasma-spot region as an increasing \ce{O}~K$\alpha$ peak (Figure~\ref{fig:Ni-oxidation-comparison-gases}b, left).
This observation is different from the oxidation \emph{outside} the plasma spots observed for a flat \ce{Cu} sample (Figure~\ref{fig:Cu-oxidation-CO2}).
This may be caused by a more pronounced sputtering of \ce{Cu} compared to \ce{Ni}, where any oxidized \ce{Cu} in the central plasma spot is directly removed by ion bombardment.
In addition, the ion dose applied to the \ce{Ni} nanoparticles (Figure~\ref{fig:Ni-oxidation-comparison-gases}, \SI{5}{\uA}) was lower than for bare \ce{Cu} (Figure~\ref{fig:Cu-oxidation-CO2}, about \SIrange{70}{120}{\uA}), resulting in more sputtering for the latter.
Besides oxidation, the sputtering during \ce{CO2} plasma changed the morphology of the \ce{Ni} particles inside the plasma spot from round shapes toward conical shapes, as discussed earlier (Figure~\ref{fig:Cone-formation}).
Overall, the EDS signals for \ce{Ni} and \ce{Cu} (from the underlying substrate) are nearly unchanged for \ce{CO2} plasma for this ion dose (Figure~\ref{fig:Ni-oxidation-comparison-gases}b, right).

Besides using \ce{CO2}, oxidation and sputtering of \ce{Ni} nanoparticles was also studied for a \SI{25}{\percent}\,\ce{O2}-\SI{75}{\percent}\,\ce{Ar} gas mixture (denoted as \ce{Ar}/\ce{O2} in the following).
The plasma parameters were kept the same as for \ce{CO2} (gap distance of \SI{250}{\um} and an approximate discharge current of \SI{5}{\uA}).
The oxidation of the \ce{Ni} particles by \ce{Ar}/\ce{O2} plasma is similar to \ce{CO2} plasma; the oxidation is localized to the plasma region (Figure~\ref{fig:Ni-oxidation-comparison-gases}a, right column), and the oxidation gradually increases with plasma duration (see \ce{O}~K$\alpha$ signal in Figure~\ref{fig:Ni-oxidation-comparison-gases}c, left).
It is noteworthy, that the oxygen-rich spot at \SI{0}{\s} in Figure~\ref{fig:Ni-oxidation-comparison-gases}a for \ce{Ar}/\ce{O2} (marked with a dashed arrow) results from a previous experiment.
Overall, the sputter rate of \ce{Ni} particles for \ce{Ar}/\ce{O2} plasma is higher than for \ce{CO2}.
The enhanced sputter yield for \ce{Ar}/\ce{O2} plasma is evident from the change in \ce{Ni} and \ce{Cu}~K$\alpha$ signals in the right plot in Figure~\ref{fig:Ni-oxidation-comparison-gases}c, where the \ce{Ni}/\ce{Cu} signal decreases/increases due to the continuous removal of \ce{Ni} particles and subsequent exposure of the underlying \ce{Cu} support.
This aspect is also slightly visible as a reduction of \ce{O} signal in the central part of the plasma spot in the \ce{O} elemental map after \SI{60}{\s} (Figure~\ref{fig:Ni-oxidation-comparison-gases}a).
After the removal of the oxidized \ce{Ni} particles in this area, the underlying \ce{Cu} support is not oxidized \emph{inside} the plasma-spot region, leading to the observed \ce{O} depletion (cf.~with \ce{O} maps in Figure~\ref{fig:Cu-oxidation-CO2}f).
This observation qualitatively agrees with simulated sputter yields using SRIM  (Table~\ref{tab:sputter-yields-srim} in the supplementary information), where \ce{Ar} has higher sputter yields $Y$ than \ce{O}.
However, \ce{CO} is another typical molecule in \ce{CO2} plasmas~\cite{willemsMassSpectrometryNeutrals2020} that could not be simulated and compared with \ce{Ar} using SRIM.

\begin{figure*}[tb]
    \centering
        
    \includegraphics[width=\linewidth]{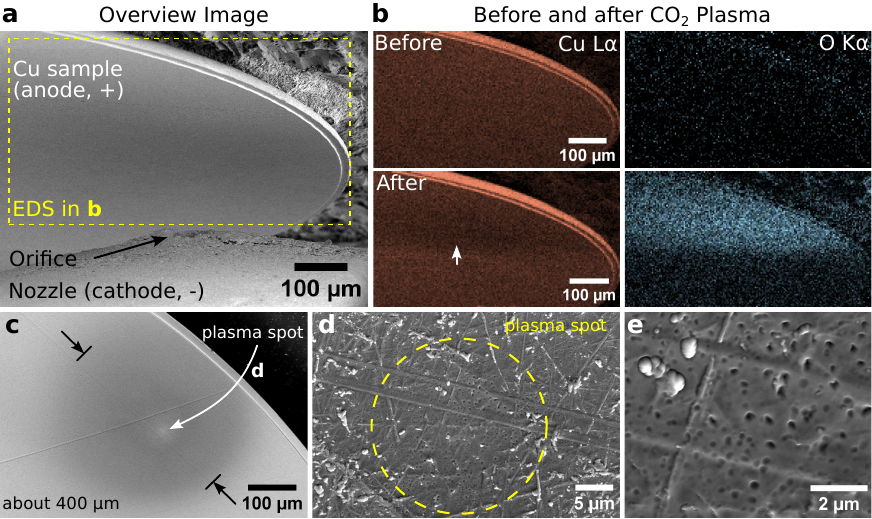}
    \caption{Local oxidation of a polished \ce{Cu} surface under \ce{CO2} plasma treatment with reversed electrode polarity. \textbf{a}~Overview SEM image of the plasma gap with the EDS region for~(b) marked with a dashed rectangle. Note the reversed nozzle/sample polarities. \textbf{b}~Quasi \emph{in-situ} EDS measurements before (upper row) and after (lower row) \ce{CO2} plasma treatment. The increased \ce{O} signal is caused by oxidation and re-deposition of oxidized \ce{Cu} from the nozzle. \textbf{c}~BSE-SEM image (\SI{5}{\keV}) of the plasma treated after plasma experiments. \textbf{d}~Higher magnification SE-SEM image of the central plasma spot. \textbf{e}~Pits formed in the central plasma spot, probably caused by ion$^{-}$ sputtering.}
    \label{fig:RevPol-Oxidation-CO2-Cu}
        
\end{figure*}

Since the sputtering is primarily caused by the bombardment of the grounded sample surface (relative to a positively biased nozzle) with positively charged ions, switching the polarity between the nozzle and the sample can mitigate sputtering.
This aspect was verified experimentally by switching the polarity upon using another DC-DC converter (XP Power, CA12N) than the previously used one (XP Power, CA20P).
The experiment was then repeated using again \ce{CO2} gas and a \ce{Cu} target.
The experimental setup is shown in Figure~\ref{fig:RevPol-Oxidation-CO2-Cu}a with the EDS acquisition area marked with a dashed line. 
The polarity between the nozzle and the sample is reversed compared to all other conducted measurements in this work.
Comparison of the \ce{O} elemental maps before and after plasma treatment (Figure~\ref{fig:RevPol-Oxidation-CO2-Cu}b) reveals a pronounced oxidation of the surface in a comparatively wide area (about \SI{400}{\um} diameter), i.e., larger than the actual plasma spot.
The latter is not clearly visible in the highly tilted view onto the \ce{Cu} target's surface in Figure~\ref{fig:RevPol-Oxidation-CO2-Cu}b, but it is visible in the top-view BSE-SEM image in Figure~\ref{fig:RevPol-Oxidation-CO2-Cu}c.
This BSE-SEM image was captured during the investigation of the same sample after the plasma experiments using standard SEM imaging parameters.
The top-view BSE-SEM image in Figure~\ref{fig:RevPol-Oxidation-CO2-Cu}c reveals the plasma spot with a higher image intensity relative to the surrounding dark area related to the oxidized \ce{Cu} surface.
Note that a low primary electron energy of \SI{5}{\keV} was used for BSE imaging to increase surface sensitivity.
The increased BSE-image intensity of the bright plasma spot (Figure~\ref{fig:RevPol-Oxidation-CO2-Cu}c) can be explained by mild sputtering in this region by negatively charged ions bombarding the positively charged \ce{Cu} surface.
This removes the oxide layer and reveals metallic \ce{Cu}, ultimately leading to higher BSE image intensity due to a higher average $Z$ than the surrounding oxidized \ce{Cu} surface.
Even though mild sputtering is present, no large pit or hole is visible in the plasma-spot region (Figures~\ref{fig:RevPol-Oxidation-CO2-Cu}d and e) compared to the initially used negative sample polarity (Figure~\ref{fig:Cu-oxidation-CO2}d).
The plasma spot area has a diameter of about \SI{25}{\um} (marked with a dashed circle in Figure~\ref{fig:RevPol-Oxidation-CO2-Cu}d) and shows the formation of small pits with \SIrange{200}{300}{\nm} (surface) diameter (Figure~\ref{fig:RevPol-Oxidation-CO2-Cu}e).
These pits are likely caused by the sputtering process and may show its initial stage.
Overall, the sputtering of the sample surface is highly reduced when the sample surface is positively biased relative to the nozzle.\newline
In the configuration shown in Figure~\ref{fig:RevPol-Oxidation-CO2-Cu}a, the mainly positively charged ions are accelerated toward the negatively biased nozzle, resulting in sputtering of the nozzle surface.
Indeed, the orifice diameter increased after these experiments and sputtered material was re-deposited inside the orifice (Figure~\ref{fig:suppl:Nozzles}).
The sputtered nozzle material is likely also re-deposited onto the opposing sample surface.
Since the same nozzle was used throughout all experiments here, previously deposited sample material (mostly \ce{Cu}) \emph{onto} the nozzle from earlier experiments is now sputtered and re-deposited \emph{from} the nozzle onto the sample (see the schematic in Figure~\ref{fig:suppl:Nozzles}j).
In our case, the orifice area is mostly covered with (oxidized) \ce{Cu} (Figure~\ref{fig:suppl:Nozzles}b) before the experiments shown in Figure~\ref{fig:RevPol-Oxidation-CO2-Cu}, meaning that part of the oxidized region is likely caused by redeposited \ce{Cu} oxide from the nozzle.\newline
To test this hypothesis, the gas was switched from \ce{CO2} to \ce{N2}, and the plasma-treated area showed a \ce{N} and \ce{O} signal (see Figure~\ref{fig:suppl:RedepositionN2} in the supplementary information).
For a \ce{N2} plasma on a \ce{Cu} surface, an \ce{O} signal is unexpected and should not be present without considering the aforementioned re-deposition effects.
Our results suggest that part of the \ce{O} signal in Figure~\ref{fig:RevPol-Oxidation-CO2-Cu}b is caused by re-deposited oxidized \ce{Cu} from the nozzle from previous experiments.
Even though this effect is undesired for pure oxidation with plasma-generated radicals, it may be interesting to study film growth during sputtering.

In summary, oxidation of \ce{Ni} nanoparticles was observed for \ce{CO2} and \ce{Ar}/\ce{O2}. 
Oxidation is limited to the central plasma-spot region.
For the same ion dose, \ce{Ar}/\ce{O2} sputtering of \ce{Ni} nanoparticles is more pronounced than for \ce{CO2}.
In contrast, oxidation of a flat \ce{Cu} surface occurs around the central plasma-spot region, which is mostly sputtered rather than oxidized.
Sputtering with positively charged ions causes rapid removal of sample material when the nozzle is used as an anode (positive polarity).
This results in pits and holes in the central plasma region.\newline
Sputtering of the sample can be strongly reduced by reversing the polarity between the sample and the nozzle, leading to less damage during oxidation.
However, sputtering of the nozzle material in this configuration causes damage to the tip of the nozzle and redeposition of this material onto the sample surface.
Pure sample oxidation without sputtering or redeposition of material requires other plasma configurations.

\subsection{Limitations and outlook}

The current setup presented here demonstrates significant advances compared to the state-of-the-art, including a stable DC discharge and true \emph{in-situ} SEM imaging. 
This enables further research regarding plasma-surface interactions, plasma physics, sputtering, and more. 
However, certain limitations remain, particularly in terms of expanding the scope of potential research areas.
For example, fields such as plasma catalysis or biomedical applications of plasma are growing rapidly, increasing the need for more advanced experimental techniques to study, e.g., plasma-catalyst or plasma-cell interactions~\cite{bogaerts2020PlasmaCatalysis2020, vonwoedtkePlasmasMedicine2013}. 
For such research topics, this setup is currently unsuited since sputtering of the sample (or redeposition of material from the nozzle) is undesirable and prevents studying the samples under relevant conditions. 
In order to study such samples, the sputtering behavior of the plasma should be eliminated.
In principle, the current setup could be optimized further to reduce the discharge voltage to decrease the ion energy, lowering the sputtering rates. 
One potential approach would be to increase the ballast resistor in the system, to limit the current and lower the discharge voltage. 
Another approach would be to further increase the pressure, as it is expected that the current setup operates below the optimum value. 
However, increasing the gas flow rate would require an upgrade to the pumping system of the SEM since the current experiments were performed at the limit of the microscope when operating in high-vacuum mode. 
The pressure could also be increased by decreasing the gap distance, but this would then also increase the probability of unwanted arcing behavior, as was also observed in our experiments. \newline
Rather, we believe that in order to expand the research potential of (quasi) \emph{in-situ} plasma in SEM experiments without sputtering or redeposition effects, a fundamentally different plasma type may be required. 
However, this would require a significant alteration of the plasma setup and a complete redesign of the electronics. 
A number of plasma types could be of interest, each with their potential applications and limitations, as well as practical drawbacks. \newline
A common plasma discharge is the dielectric barrier discharge (DBD)~\cite{fridmanPlasmaPhysicsEngineering2011}. 
This alternating current (AC, or pulsed) discharge is characterized by a dielectric layer covering one or both electrodes, limiting the current and thus preventing arc formation. 
This is a non-thermal plasma which is often used in plasma catalysis and biomedical research. 
However, DBD plasmas are generally filamentary, where the filaments consist of microdischarges (short duration, high current discharges). 
These filaments make the plasma treatment of the sample heterogeneous, complicating the analysis, and cause issues with electromagnetic interference. 
In principle, DBDs can be operated in a uniform mode~\cite{massinesRecentAdvancesUnderstanding2009}, but this requires precise tuning of all relevant parameters (including the dielectric material, voltage, frequency, discharge gas, and pressure) further impeding rapid development of such an experimental setup. \newline
An alternative discharge based on the DBD is the so-called surface discharge. 
This plasma is similar to the DBD, but one of the electrodes is embedded or below the dielectric, whereas the other electrode is placed on the surface of the dielectric. 
With this, the discharge will be generated at the surface of the dielectric. 
This plasma still requires AC or pulsed power, but is generally more convenient to operate in a uniform mode~\cite{fridmanPlasmaPhysicsEngineering2011}.\newline
Another approach could be using a plasma jet. 
Many geometries exist, either powered by DC, pulsed, or AC power, but they all have in common that the plasma is generated within a device, after which it flows outwards, e.g., to a sample~\cite{luAtmosphericpressureNonequilibriumPlasma2012}. 
The main difference with the setup presented here is that in the current setup, the plasma is generated in the gap between the nozzle and the sample rather than in the nozzle and sent to the sample. 
A main advantage of such a plasma jet could be the elimination of the sputtering behavior, as charged plasma species are not predominant (or even absent in the so-called afterglow). 
Based on this geometry, an electron beam plasma can be generated~\cite{fridmanPlasmaPhysicsEngineering2011}, of which a variation was previously introduced in an SEM~\cite{muldersInsituLowEnergy2016a}. 
In such plasmas, a high-energy electron beam is sent through a neutral gas, where the electrons ionize gas molecules. 
The plasma can then be sent to a sample through a gas flow, or the ions/electrons could be selectively attracted by biasing the sample. 
An external AC or DC circuit can also be added to further sustain and alter the plasma discharge, depending on the desired properties. 
Having access to a high-energy electron beam makes an SEM promising to further explore such plasmas. \newline
Note that all AC or pulsed-powered plasmas are very likely to interfere with the true \emph{in-situ} imaging of the SEM since the electron beam will be deflected periodically during scanning, drastically decreasing the image resolution. 
Depending on the desired experiment, this issue could be overcome by turning off the plasma during image acquisition, though this does limit the \emph{in-situ} capabilities of the setup.\newline
Further, introducing a microplasma may enable very different experiments and applications.
On the one hand, the \emph{in-situ} plasma may lead to new analytical techniques in an SEM, such as glow discharge optical emission spectroscopy (GDOES)~\cite{grimmNeueGlimmentladungslampeFuer1968, gamezSurfaceElementalMapping2012}, where the emission from sputtered material in a plasma is studied while ablating the sample material for depth profiling (similar to secondary ion mass spectroscopy in focused ion beam instruments~\cite{pillatschFIBSIMSReviewSecondary2019}).
On the other hand, established (e.g., EDS or wavelength dispersive x-ray spectroscopy, WDS~\cite{llovetElectronProbeMicroanalysis2021}) or more recently available (e.g., electron energy loss spectroscopy, EELS~\cite{broduschElectronEnergylossSpectroscopy2019}) analytical methods in SEMs may have the potential to probe the ionic species in the plasma cloud. 
This would provide essential and direct \emph{in-situ} feedback for plasma simulation codes and holds promise for improved control over plasma setups. 

\section{Conclusions} \label{conclusions}

A custom-built microplasma setup was realized inside an SEM based on the design by \citet{matraLocalSputterEtching2013}.
A nozzle with a small orifice feeds a gas into the evacuated SEM chamber, from which a plasma can be generated by applying a certain electrical potential.
Stable DC glow discharge plasmas with \ce{Ar}, \ce{Ar/O2}, \ce{CO2}, and \ce{N2} gases were successfully generated in the SEM's vacuum chamber.
In general, larger discharge currents were measured for higher gas flow rates and smaller gap distances.
A non-uniform gas-pressure profile was observed in the plasma gap, which --- in combination with a non-uniform electric field of the electrode geometry --- complicates a direct comparison of the shown setup with conventional plasma systems.
Simultaneous SEM imaging with SEs and BSEs during plasma operation was demonstrated, enabling \emph{in-situ} studies of sample-plasma interactions in the SEM.\newline
A few exemplary plasma-sample interactions were studied. 
Sputtering of \ce{Cu} surfaces and \ce{Ni} nanoparticles under different gases was observed. 
The lower sputter yield of the \ce{Ni} particles compared to the \ce{Cu} support, as well as the incidence-angle dependence of the sputter yield, results in the local formation of cones in the plasma-treated area.
The same phenomenon was studied with conventional plasma reactors, which shows that our setup can replicate such experimental conditions on the local scale of several tens of \unit{\um}.
Local oxidation of \ce{Cu} and \ce{Ni} was observed for \ce{CO2} gas and an \ce{Ar}/\ce{O2} gas mixture.
At the same time, however, the sample was either simultaneously sputtered away by ion bombardment on the sample, or nozzle material was redeposited on the sample by sputtering of the nozzle.
These limitations might be overcome by further optimizations of the setup, though for applications where sputtering is detrimental, other types of plasma are to be considered.\newline
In conclusion, we have demonstrated that \emph{in-situ} studies of plasma-sample interactions in a modern SEM are possible.
This approach provides direct insight into morphological and chemical changes (via EDS) of the sample during and after plasma treatment.
Overall, this may lead to a better understanding of plasma physics and plasma-surface interactions.

\section{Experimental} \label{methods}

\subsection{SEM Operation with the Plasma Setup}

\begin{figure*}[tb]
        
    \includegraphics[width=\linewidth]{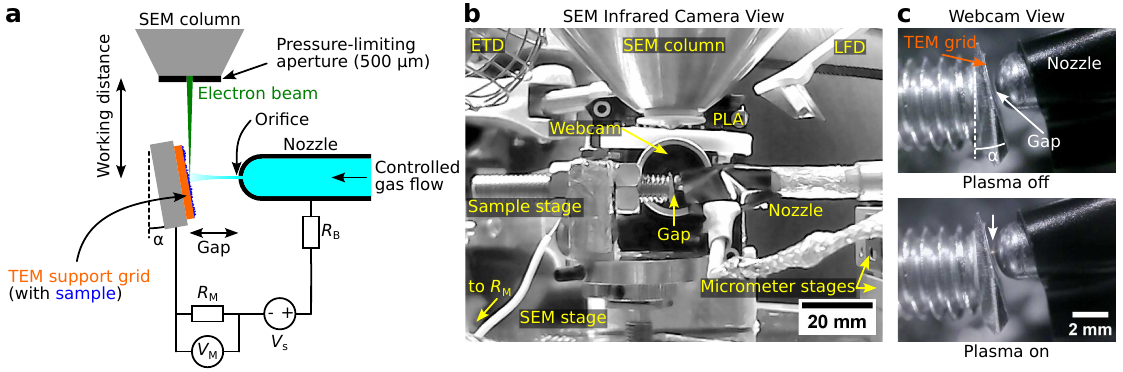}
    \caption{Schematics and images of the plasma-in-SEM setup. \textbf{a}~Schematic showing the experimental setup and the most important components. Gas flows from a nozzle orifice over an adjustable gap distance toward a sample surface. A high voltage $V_\text{S}$ is applied to ignite the plasma. The sample surface is slightly tilted at an angle $\alpha$ toward the SEM incidence, allowing for \emph{in-situ} SEM imaging. \textbf{b}~Image of the setup taken with the built-in infrared camera of the SEM showing the setup. A few additional components are shown compared to \textbf{a}, such as a webcam and the electron detectors, ETD and LFD. \textbf{c}~Higher-magnification side-view of the plasma region using the webcam without plasma (upper) and with ignited plasma (lower, the microplasma is marked with an arrow)}
    \label{fig:Plasma-Setup-Scheme}
        
\end{figure*}

Plasma experiments were performed using an FEI Quanta 250 ESEM equipped with an Oxford Instruments X-Max EDS detector (\SI{80}{\mm^2} sensor area).
Figure~\ref{fig:Plasma-Setup-Scheme}a schematically shows the main parts of the plasma setup that was built in-house.
A horizontally aligned steel nozzle with a small orifice (SS-1/8-TUBE-CAL-20, \SI{20}{\um} nominal orifice diameter, Lenox Laser) is fixed opposite to a nearly vertically aligned sample surface.
The sample surface is slightly tilted with an angle $\alpha \approx \SI{10}{\degree}$ toward the electron beam for better SEM imaging conditions.
The sample-nozzle distance (\enquote{Gap} in Figures~\ref{fig:Plasma-Setup-Scheme}a-c) determines the plasma gap distance and can be adjusted by moving the sample with SEM microscope stage controls.
A gas flows from the nozzle into the gap toward the sample surface.
The nozzle can be biased with a DC voltage $V_\text{S}$ in the range of \SIrange{-1.25}{2}{\kV}, i.e., with a positive or negative polarity relative to the sample.
A ballast resistance $R_\text{B} = \SI{4.3}{\mega\ohm}$ is used to limit the discharge current.
The discharge current $I_\text{D} = V_\text{M}/R_\text{M}$ is measured by the voltage drop $V_\text{M}$ across a $R_\text{M} = \SI{1}{\kilo\ohm}$ resistor.\newline
Figure~\ref{fig:Plasma-Setup-Scheme}b displays the experimental setup with an image taken with the microscope's built-in infrared (IR) camera. 
A few additional components compared to the schematic in Figure~\ref{fig:Plasma-Setup-Scheme}a are visible, which are explained from top to bottom in the following.
The ETD and the large-field detector (LFD) are used for SEM imaging in high-vacuum and low-vacuum modes, respectively.
The shown images in this work are mainly SE-SEM images.
Selected BSE-SEM images are mentioned explicitly in the text.
A pressure-limiting aperture (PLA) with a \SI{500}{\um} diameter is mounted on the SEM pole piece to restrict gas flow into the microscope column.
An IR-USB webcam (Arducam B0205) is mounted in addition to the microscope's built-in IR camera to improve imaging conditions of the plasma and control the gap distance.
The sample stage consists of a threaded metal rod that is rigidly fixed with two nuts to a Teflon piece.
The Teflon piece isolates the sample from the microscope stage to prevent current flow through the latter and possible damage to the microscope.
Instead, the current flows via a cable to the measurement resistor $R_\text{M}$.
The sample stage with the threaded metal rod and the Teflon block are fixed on an SEM stub, which itself is fixed on the moveable SEM stage.
Two micrometer stages (Thorlabs MS3/M) are used to laterally position the nozzle close to the optical axis (below the SEM pole piece) before closing the SEM chamber.
The nozzle and the webcam are mounted on a \ce{Al} platform that is fixed above the moving microscope stage.
The height of the \ce{Al} platform can be adjusted to change the working distance between the SEM column and the sample (typically \SI{15}{\mm}).
The gas line and electrical connections are routed through a custom home-made feedthrough flange.
\newline
A detailed image of the plasma gap is shown in the webcam view (Figure~\ref{fig:Plasma-Setup-Scheme}c).
Commercially available grids or apertures made for transmission electron microscopy (TEM) with \SI{3}{\mm} diameter (Gilder Grids GA50 \ce{Cu} apertures) were typically used as sample or sample support for nanoparticles.
The sample is mounted on an \ce{Al} wedge with conductive \ce{Ag} paste (EM-Tec AG15). 
The \ce{Al} wedge was ground at an angle $\alpha$ and fixed to the threaded metal rod's end with conductive \ce{Ag} paste.
The lower image in Figure~\ref{fig:Plasma-Setup-Scheme}c shows the working setup with a glowing DC microplasma.
More details about the experimental setup can be found in the supplementary information (Figure~\ref{fig:suppl:SEM-Setup}).

\subsection{Plasma Operation}

Plasma experiments were performed in the high-vacuum mode of the microscope since undesired discharges in the SEM chamber in low-vacuum mode were observed when applying high voltage between the nozzle and the sample.
The high-vacuum mode reached a stable chamber pressure of around \SI{2e-2}{\Pa} while providing a gas flow of about \SIrange{2}{8}{\sccm} through the nozzle (\SI{20}{\um} nominal orifice diameter as per the manufacturer) into the microscope chamber.
The gas flow was monitored using an Alicat flow meter (M-200SCCM-D/5M).
We used \ce{CO2} (purity \SI{99.995}{\percent}), \ce{Ar} (\SI{99.9999}{\percent}), and \ce{N2} (\SI{99.9999}{\percent}) gases, and a \SI{75}{\percent}\ce{Ar}/\SI{25}{\percent}\ce{O2} gas mixture (measured: \SI{74.88}{\percent}/\SI{25.12}{\percent}) in this work (bought from Air Products). \newline 
The plasma was operated by applying and controlling the voltage difference on the nozzle relative to the sample.
A DC-DC converter with a \SI{1}{\mega\ohm} output resistor (CA20P or CA12N depending on polarity, XP Power) was powered by an RS~PRO IPS-3303 power supply.
The \SI{1}{\mega\ohm} output resistor limits the output current of the DC-DC converter in standalone usage for user safety.
The output resistor is in series with a \SI{3.3}{\mega\ohm} resistor, resulting in a  total ballast resistance $R_\text{B} = \SI{4.3}{\mega\ohm}$. 
The output high voltage $V_\text{S}$ of the DC-DC converter was adjusted with a control voltage between \SIrange{0}{5}{\V} using a Keysight E36106B power supply.
After plasma ignition, the discharge current was regulated by adjusting $V_\text{S}$ with the control voltage.
Voltage-current characteristics of the plasma were measured with a Keithley 2400 source measurement unit. 
The highest source voltage of \SI{2}{\kV} was applied, after which the source voltage was gradually reduced while registering the current until no discharge current was measurable.
The discharge voltage of the DC plasma $V_\text{D}$ is calculated as $V_\text{D} = V_\text{S} - I_\text{D}\left(R_\text{B} + R_\text{M}\right)$~\cite{gudmundssonFoundationsDCPlasma2017}.

\subsection{Sample Preparation}

A \ce{Cu} TEM aperture (\SI{50}{\um}, Gilder Grids GA50) with a diameter of \SI{3}{\mm} and a thickness of about \SI{30}{\um} was used in most experiments to ensure a well-defined, flat electrode opposing the nozzle.
For experiments with nanoparticles, commercial \ce{Ni} particles (nanopowder, \SI{<100}{\nm} nominal average particle size, \SI{>99}{\percent} purity, Sigma-Aldrich, CAS number 7440-02-0) were mixed with acetone and then drop cast on the \ce{Cu} disc.
After solvent evaporation, a thin film of \ce{Ni} particles is left on the \ce{Cu} surface.
Drop casting was repeated multiple times until the TEM aperture was fully covered with \ce{Ni} particles.

\subsection{Data Processing}
\emph{Fiji}~\cite{schindelinFijiOpensourcePlatform2012} was used for general image processing.
Images were stitched together using the \enquote{Grid/collection stitching} plugin~\cite{preibischGloballyOptimalStitching2009}. 
Image series were registered using the \enquote{Descriptor-based series registration (2d/3d + t)} plugin~\cite{preibischSoftwareBeadbasedRegistration2010}.
The background-corrected x-ray peak intensities (net intensities) for the EDS maps were extracted using the \enquote{TruMap} function in the Oxford Instruments \emph{AZtec} software (version~2.1).
Additional analyses of extracted (summed-up) EDS spectra from specific regions were processed with the \emph{HyperSpy} Python package~\cite{penaHyperspyHyperspyRelease2022}.

\section{Data Availability Statement}
Raw data files and data-treatment scripts are available at Zenodo~\cite{grunewaldSupplementaryInformationInsitu2023} (\url{https://doi.org/10.5281/zenodo.8042029}).\newline

\section{Author Contributions}
\textbf{LG}: Conceptualization, Methodology, Investigation, Software, Validation, Formal Analysis, Data Curation, Visualization, Writing -- Original Draft. 
\textbf{DC}: Conceptualization, Methodology, Investigation, Writing –- Review \& Editing. 
\textbf{RDM}: Conceptualization, Methodology, Investigation, Validation, Writing -- Original Draft.
\textbf{AO}: Conceptualization, Methodology, Writing –- Review \& Editing.
\textbf{SVA}: Conceptualization, Supervision, Project Administration, Funding Acquisition, Writing –- Review \& Editing.
\textbf{AB}: Conceptualization, Supervision, Project Administration, Funding Acquisition, Writing –- Review \& Editing.
\textbf{SB}: Conceptualization, Supervision, Project Administration, Funding Acquisition, Writing –- Review \& Editing.
\textbf{JV}: Conceptualization, Methodology, Supervision, Project Administration, Funding Acquisition, Writing –- Review \& Editing

\section{Conflicts of Interest}
There are no conflicts to declare.

\section{Acknowledgments}
LG, SB, and JV acknowledge support from the iBOF-21-085 PERsist research fund.
DC, SVA, and JV acknowledge funding from a TOPBOF project of the University of Antwerp (FFB 170366).
RDM, AB, and JV acknowledge funding from the Methusalem project of the University of Antwerp (FFB 15001A, FFB 15001C).
AO and JV acknowledge funding from the Research Foundation Flanders (FWO, Belgium) project SBO S000121N.

\begin{suppinfo}

Details of $V$-$I$ measurements, exemplary Paschen curve for \ce{N2}, sputter yield simulations, details about EDS spectrum comparison, SEM/EDS characterization of the orifice and redepostion effects, and more details about the experimental setup are found in the supplementary information.

\end{suppinfo}

\clearpage
\FloatBarrier

\setcounter{figure}{0}
\renewcommand\thefigure{S\arabic{figure}}

\setcounter{table}{0}
\renewcommand\thetable{S\arabic{table}}

\twocolumn[\section{Supplementary Information for \enquote{\emph{In-situ} Plasma Studies using a Direct Current Microplasma in a Scanning Electron Microscope}}]

\subsection{Details about Voltage-Current-Characteristic Measurements}
\begin{figure*}[tb]
        
    \includegraphics[width=\linewidth]{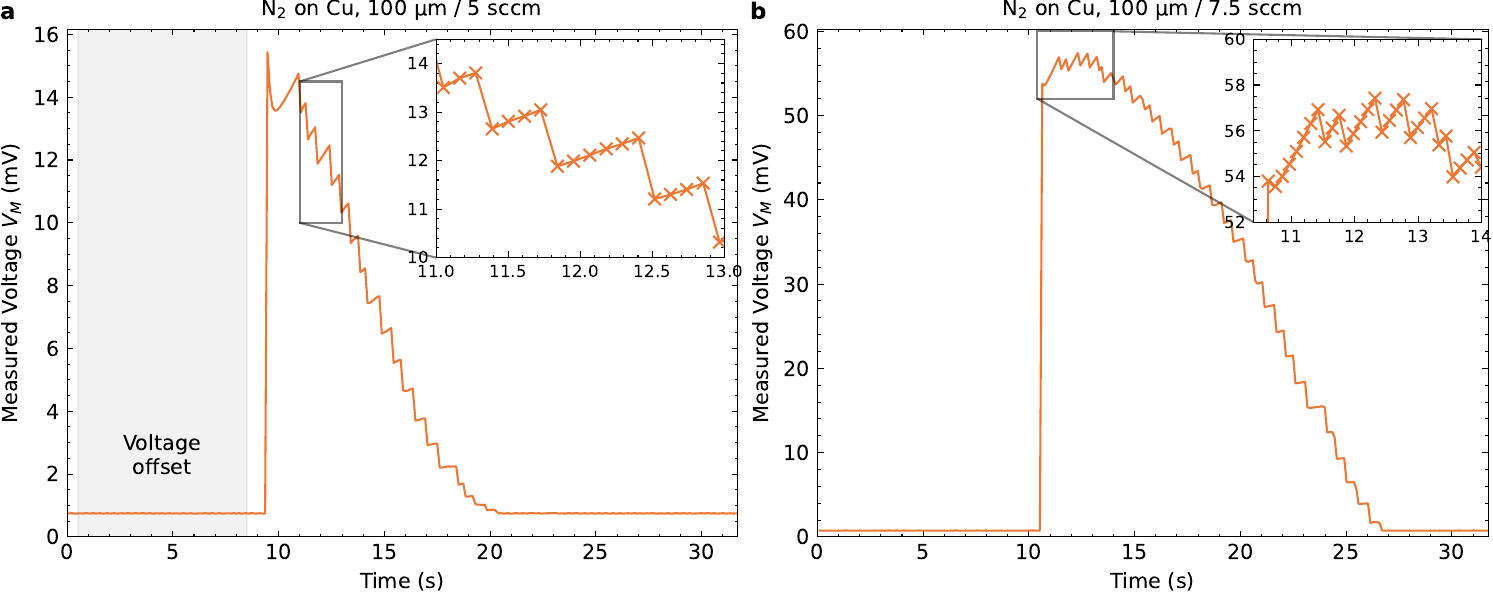}
    \caption{Measurement problems of voltage-current characteristics, shown for two different measurements. \textbf{a}~The shaded area marks a reference region without actual microplasma operation, which was used to determine and subtract the voltage offset. The sharp onset shows the plasma ignition for the highest source voltage of \SI{2}{\kV}, followed by subsequent reduction in \SI{40}{\V} steps. The ideally-horizontal steps show a positive slope due to a continuous discharge current increase, especially in the first few seconds of plasma operation. This is attributed to rapid sputtering of the sample surface. \textbf{b}~Another example of higher gas flow rate, resulting in higher discharge currents. The sputtering is faster, and the slope is more pronounced than in (a).}
    \label{fig:suppl:SI-Problems-VI-Measurements}
        
\end{figure*}

The main text shows voltage-current characteristics of the generated microplasma in the scanning electron microscope's (SEM's) chamber.
A problem during measurements was the continuous sputtering of the sample surface when using the nozzle as an anode with positive bias.
Typical measurements of the voltage drop across the measurement resistor $R_\text{M}$ versus the measurement duration are shown in Figure~\ref{fig:suppl:SI-Problems-VI-Measurements}.
All measurements were started by first applying the highest possible source voltage $V_\text{S}$ with the DC-DC converter (\SI{2}{kV}, visible as strong onset in the plots) and then gradually decreasing the source voltage with \SI{40}{V} steps until no voltage across the measurement resistor $R_\text{M}$ was measurable anymore.
A small parasitic offset voltage was measured and subtracted from a reference region (e.g., the shaded area in Figure~\ref{fig:suppl:SI-Problems-VI-Measurements}a).
Ideally, the voltage steps in the measured curve should be horizontal plateaus, whereby each step corresponds to a defined step in the applied source voltage (here in steps of \SI{40}{\V}).
However, as visible in the inset in Figure~\ref{fig:suppl:SI-Problems-VI-Measurements}a, the discharge current was not stable but instead steadily increasing, especially in the first few seconds of plasma operation and at high currents. 
We attribute this to surface sputtering and rapid change of the electrode geometry, which significantly decreased the resistance of the gap and thus increased the discharge current (given the constant applied voltage).
Even though it is, in principle, possible to correct the slope of the $V_\text{M}$-time curve, we opted to simply calculate the average value of each voltage step by manual extraction (see Jupyter notebook on Zenodo~\cite{grunewaldSupplementaryInformationInsitu2023}).
Figure~\ref{fig:suppl:SI-Problems-VI-Measurements}b shows a more extreme example of higher discharge currents, resulting in faster surface sputtering and an even more pronounced discharge-current increase over time.
A strong slope is visible in the inset figure for the first few seconds of plasma operation. 
The voltage steps become horizontal at around \SI{20}{\s} in the plot.

\begin{figure}[htb]
        
    \includegraphics[width=\linewidth]{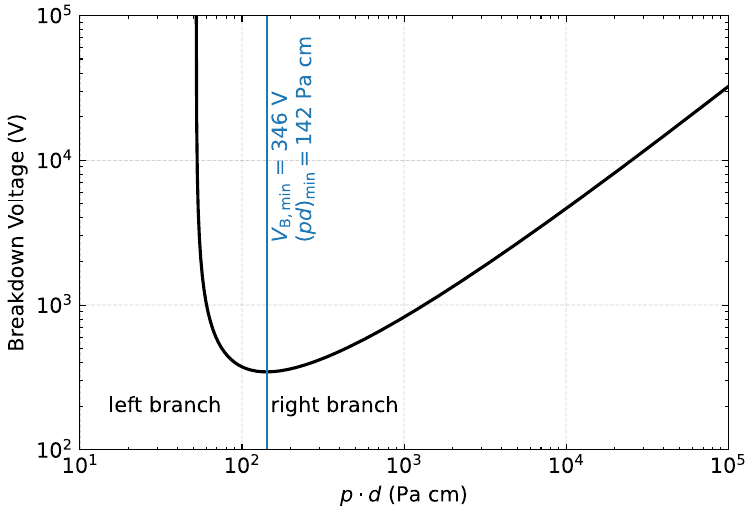}
    \caption{A calculated Paschen curve for \ce{N2} to show the two typical plasma regimes for DC glow discharges on the left and right branch relative to the minimum of the curve (referred to as \enquote{optimum} in the main text).}
    \label{fig:suppl:SI-PaschenCurve-Nitrogen}
        
\end{figure}

A theoretical Paschen curve calculated for \ce{N2} is shown in Figure~\ref{fig:suppl:SI-PaschenCurve-Nitrogen}.
The vertical line marks the minimum breakdown voltage $V_\text{B,min} = \SI{346}{\V}$ of the curve at $(pd)_\text{min} = \SI{142}{\Pa\,\cm}$.
The breakdown voltages were calculated according to the equation~\cite{gudmundssonFoundationsDCPlasma2017}
\begin{align}
{\displaystyle V_{\text{B}}={\frac {Bpd}{\ln(Apd)-\ln \left[\ln \left(1+{\frac {1}{\gamma _{\text{see}}}}\right)\right]}}}\quad\text{,}
\end{align}
with the parameters $A = \SI{11.8}{\cm^{-1}\,\torr^{-1}} = \SI{8.85e-2}{\cm^{-1}\,\Pa^{-1}}$, $B = \SI{325}{\V\,\cm^{-1}\,\torr^{-1}} = \SI{2.44}{\V\,\cm^{-1}\,\Pa^{-1}}$, and $\gamma_\text{see} = 0.01$ (secondary-electron-emission coefficient) for \ce{N2}~\cite{liebermanPrinciplesPlasmaDischarges2005}.
The Python code of the Wikipedia user \enquote{Krishnavedala} (\url{https://commons.wikimedia.org/wiki/File:Paschen_curves.svg}) was used and modified for Figure~\ref{fig:suppl:SI-PaschenCurve-Nitrogen}.

\begin{figure*}[htb]
        
    \includegraphics[width=\linewidth]{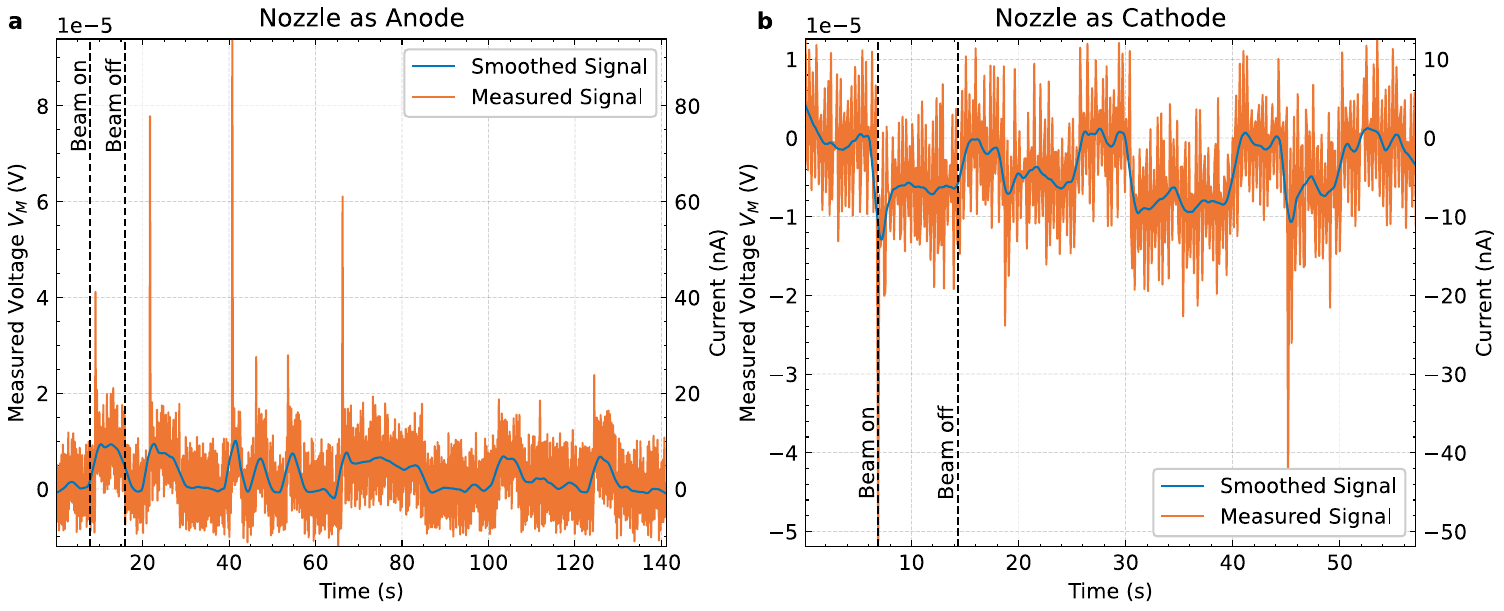}
    \caption{Measured voltage across a \SI{1}{\kilo\ohm} resistor (left abscissa, note the \num{1e-5} factor) and corresponding discharge current (right abscissa) for experimental conditions that did not result in a continuous plasma for two different nozzle-sample polarities. Raw and smoothed signals are displayed. Each step corresponds to the time the electron beam was on and scanning. The measured current of a few \unit{nA} is within the range of typical electron-beam currents for the used beam parameters.~\textbf{a}~Nozzle as the anode, and (\textbf{b})~ nozzle as the cathode.}
    \label{fig:suppl:SI-BeamCurrentPlasma}
        
\end{figure*}

No continuous plasma discharges were observed for specific combinations of (large) gap distances, (low) gas flow rates, and (low) source voltages.
In such cases, the electron beam may be able to ignite the plasma and initiate a continuous discharge. 
In other cases, a discharge current was only measured when the electron beam was on and immediately vanished after the beam was switched off (Figure~\ref{fig:suppl:SI-BeamCurrentPlasma}).
Two examples for the latter are shown in Figures~\ref{fig:suppl:SI-BeamCurrentPlasma}a and b, where each step in the signal corresponds to the electron beam being switched on or off.
A smoothed signal is plotted as well for better visibility of the steps.
The signal was smoothed using locally weighted regression (LOWESS) with \emph{HyperSpy}~\cite{penaHyperspyHyperspyRelease2022} with \emph{smoothing\_parameter=0.03} and \emph{number\_of\_iterations=1}.
The calculated current is in the \unit{nA}-range, which is typical for SEM measurements, but was not explicitly measured here. 
This implies that the generated plasma current directly relates to the electron-beam current.
The beam conditions were \SI{15}{\keV}, \SI{30}{\um} objective aperture, and spot size \num{5}.
Interestingly, some steps show an initial current spike in the non-smooth signal (about \SIrange{40}{100}{\nA} for Figure~\ref{fig:suppl:SI-BeamCurrentPlasma}a or \SIrange{20}{50}{\nA} Figure~\ref{fig:suppl:SI-BeamCurrentPlasma}b) and then the reduction to the actual electron-beam current (presumably, but not measured).

\subsection{Sputter Yield Simulation}
\begin{figure}[htb]
        
    \includegraphics[width=\linewidth]{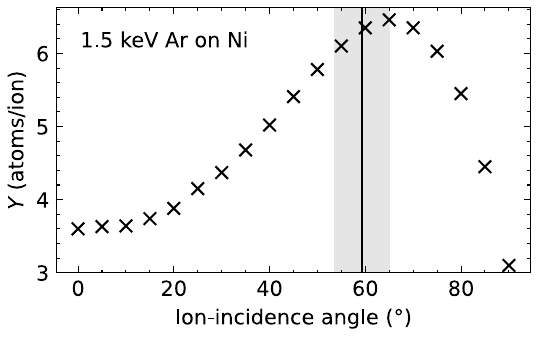}
    \caption{Simulated sputter yield $Y$ versus ion-incidence angle for \ce{Ar+} ions on a \ce{Ni} target. $Y$ increases with angle up to a maximum around \SI{65}{\degree} and then falls off again. The experimental value \SI[separate-uncertainty = true]{59.3(5.6)}{\degree} (plus/minus its standard deviation) calculated from the cone angles is marked with a vertical line (shaded area).}
    \label{fig:suppl:SRIM-Y-angle-sim}
        
\end{figure}

Monte Carlo simulations using SRIM~2013 (\url{http://www.srim.org/}) were run to investigate sputter yields for different ions and materials (Table~\ref{tab:sputter-yields-srim}) and ion-incidence angles (Figure~\ref{fig:suppl:SRIM-Y-angle-sim}).\newline
Overall, heavier ions (here \ce{Ar}) show a higher sputter yield $Y$ than lighter ions (\ce{C} and \ce{O}) on \ce{Cu} and \ce{Ni} targets.
Note, that \ce{CO2} plasma creates mostly \ce{CO} and \ce{O}~\cite{willemsMassSpectrometryNeutrals2020} instead of elemental \ce{C}, meaning that the latter is given here only for completeness. 
\ce{CO} sputtering could not be simulated with SRIM.
The higher sputter yield of \ce{Cu} compared to \ce{Ni} may explain the formation of \ce{Ni} cones on a \ce{Cu} substrate under \ce{Ar^+} bombardment.\newline
Regarding the angle dependence of the sputter yield, Figure~\ref{fig:suppl:SRIM-Y-angle-sim}, \ce{Ar+} ions with an energy of \SI{1.5}{\keV} hitting a \ce{Ni} target show that the maximum sputter yield is found for an incidence angle of about \SI{65}{\degree}.
This angle is related to the cone shape, and the measured value and its standard deviation for the maximum-sputter-yield angle are shown with a vertical line and the shaded area.
The differences between the experiment and the simulation may be explained by limited statistics for the experimental value and the simplified flat geometry used in the SRIM simulations (e.g., neglecting nanoparticle morphology) and their limited accuracy~\cite{shulgaNoteArtefactsSRIM2018}.
\begin{table}[bt]
    \caption{Sputter yields $Y$ for different \SI{1.5}{\keV} ions for normal incidence on \ce{Cu} and \ce{Ni} surfaces  obtained from Monte Carlo simulations using SRIM-2013.} 
    \label{tab:sputter-yields-srim}
    \begin{tabular}{l l l r}
    \toprule
    Ion & $Z$ & Target      & $Y$ (atoms/ion)      \\
    \midrule
    C   & 6  &  Cu  & \num{2.21}  \\
        &    &  Ni  & \num{1.83}  \\
    O   & 8  &  Cu  & \num{2.99}  \\
        &    &  Ni  & \num{2.47}  \\
    Ar  & 18 &  Cu  & \num{4.58}  \\
        &    &  Ni  & \num{3.60}  \\
    \bottomrule
    \end{tabular} 
\end{table}

\subsection{Spectrum Normalization in EDS}
\begin{figure*}[tb]
        
    \includegraphics[width=\linewidth]{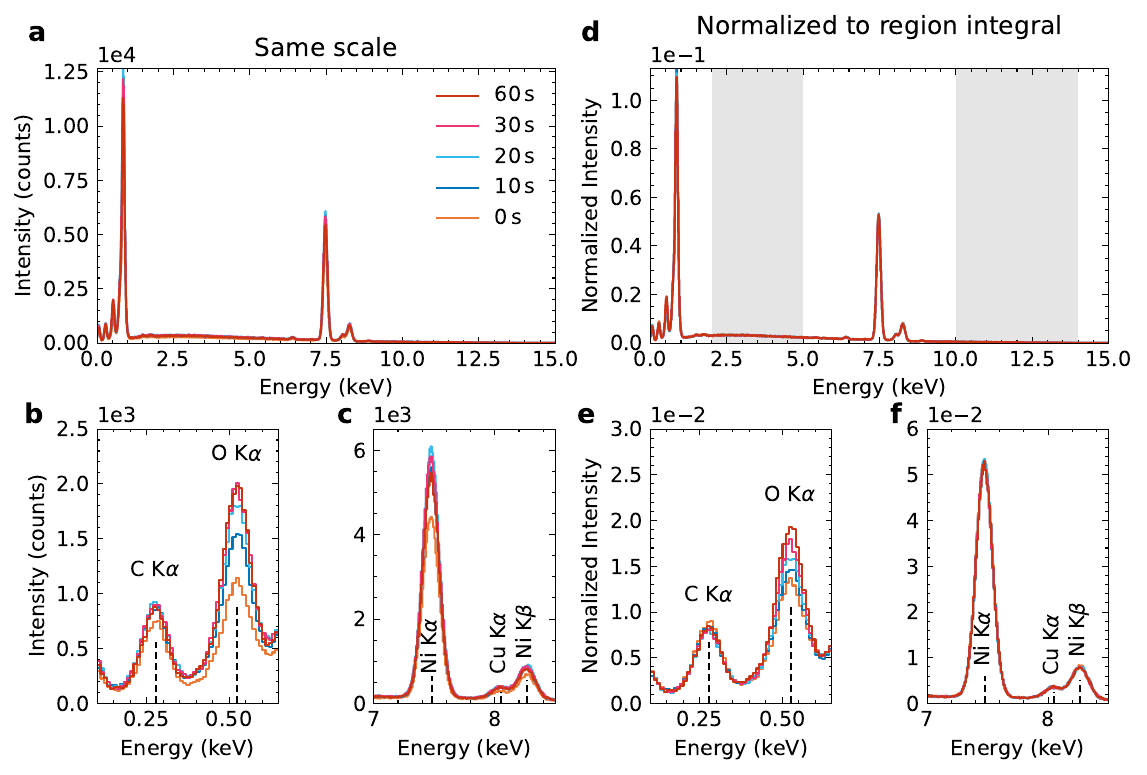}
    \caption{Comparison of EDS spectra without (left) and with (right) normalization with respect to the summed intensity in a given energy interval (shaded in gray). The upper row shows the full energy range from \SIrange{0}{15}{\keV}, and the lower row shows exemplary energy windows containing the \ce{C} and \ce{O} signal, and the \ce{Ni} and \ce{Cu} signals. The normalization accounts for varying total x-rays in the EDS spectra, impeding a direct comparison.}
    \label{fig:suppl:EDS-Normalization}
        
\end{figure*}
Figures~\ref{fig:suppl:EDS-Normalization}a--c show energy-dispersive x-ray spectroscopy (EDS) spectra without normalization and varying total electron dose, resulting in different total x-ray counts.
The upper plot shows the full energy range from \SIrange{0}{15}{\keV} (Figure~\ref{fig:suppl:EDS-Normalization}a).
The insets in the lower row (Figures~\ref{fig:suppl:EDS-Normalization}b and c) show selected energy ranges for energy windows containing the \ce{C} and \ce{O} signals (Figure~\ref{fig:suppl:EDS-Normalization}b), and the \ce{Ni} and \ce{Cu} signals (Figure~\ref{fig:suppl:EDS-Normalization}c).
For these spectra, a direct comparison is impeded by the difference in total x-ray counts, resulting in varying peak heights even without relative changes between spectra.
After normalization (Figures~\ref{fig:suppl:EDS-Normalization}d--f), the increasing \ce{O} signal is revealed (Figure~\ref{fig:suppl:EDS-Normalization}e), and the signals for \ce{Ni} and \ce{Cu} are unchanged (Figure~\ref{fig:suppl:EDS-Normalization}f).
The shaded areas in Figure~\ref{fig:suppl:EDS-Normalization}d mark the regions used for spectrum normalization.
These contain no elemental peaks and only bremsstrahlung background.
The sum of the x-ray counts in these two areas was used for normalization of the spectra.

\subsection{Characterization of the Nozzle Orifice}
\begin{figure*}[htb]
        
    \includegraphics[width=\linewidth]{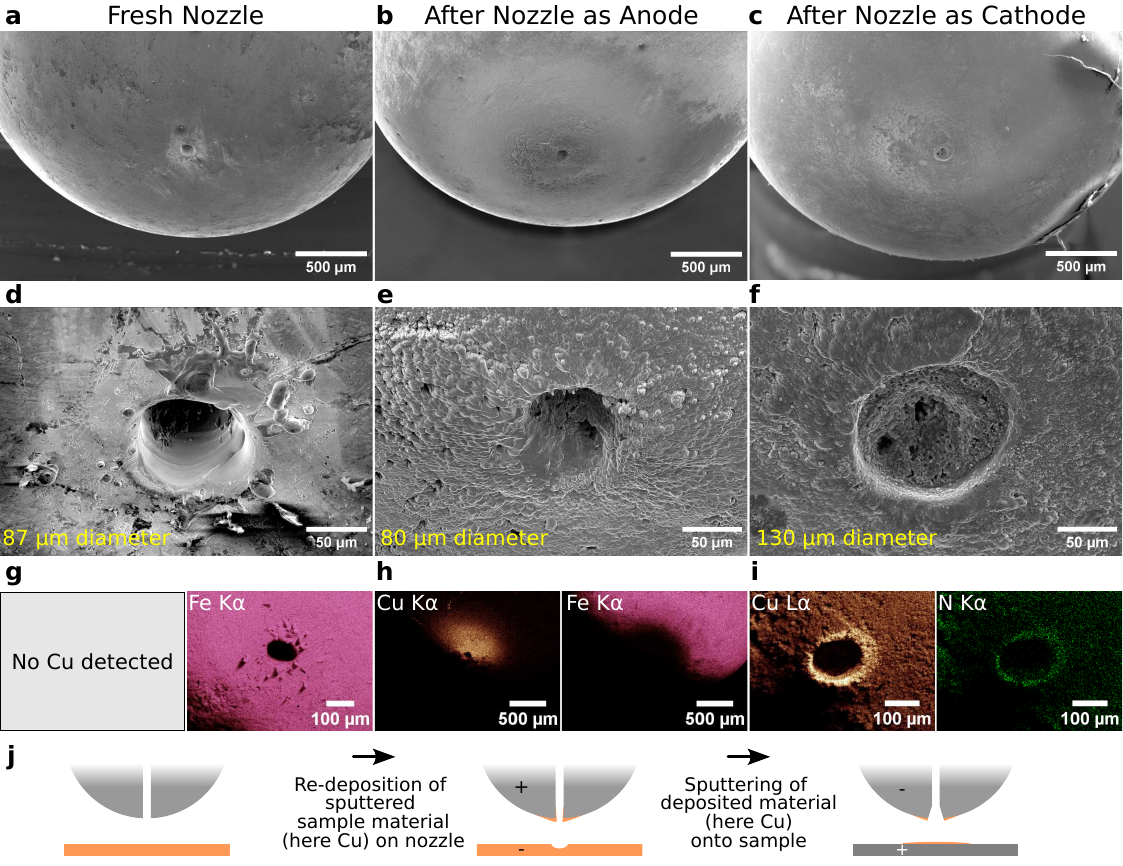}
    \caption{Low magnification (upper row) and high magnification (middle row) SEM images of the fresh (left column, below \textbf{a}) nozzle and after it was used as anode (middle column, below \textbf{b}) and cathode (right column, below \textbf{c}). EDS maps of the orifice are shown in the lower middle row starting from (\textbf{g}). Note that the absence of x-ray signals from the lower left corner in the maps is caused by the shadowing of x-rays by the nozzle. \ce{Fe} signal stems from the steel nozzle, \ce{Cu} and \ce{N} signals from material redeposition/sputtering. The last row depicts the situation of the nozzle when used as an anode or a cathode.}
    \label{fig:suppl:Nozzles}
        
\end{figure*}
SEM imaging and chemical analysis with EDS were used to inspect the nozzle's orifice between different plasma experiments (Figure~\ref{fig:suppl:Nozzles}).
The unused, fresh nozzle in the left column of Figure~\ref{fig:suppl:Nozzles} shows some contamination near the orifice (Figure~\ref{fig:suppl:Nozzles}d), but the inner walls of the laser-cut orifice are well-defined.
The outer diameter is about \SI{87}{\um}, which is substantially larger than the nominal \SI{20}{\um}.
Since the measured gas flow rate was close to the nominal value, we suspect that the orifice diameter gets gradually smaller toward the inside of the nozzle. 
The EDS elemental map of \ce{Fe}~K$\alpha$ resulting from the steel nozzle is used to show the absence of debris (Figure~\ref{fig:suppl:Nozzles}g). 
The lack of signal from the lower left corner of the EDS map is caused by shadowing effects toward the EDS detector (no direct line of sight for emerging x-rays).\newline
When the nozzle is used as an anode with a positive bias, most of the positively charged ions are hitting the sample surface, and the sputtered material is redeposited on the nozzle tip.
Since mostly \ce{Cu} apertures were used as sample material, a pronounced \ce{Cu}~K$\alpha$  signal is visible around the orifice (Figure~\ref{fig:suppl:Nozzles}h).
The deposited material reduces the outer diameter of the orifice (here to about \SI{80}{\um}).\newline
Afterward, the nozzle was used as a cathode and the ions are now mostly bombarding the nozzle instead of the sample.
Sputtering of the orifice region leads to the removal of the previously-deposited \ce{Cu}. 
The sputtered \ce{Cu} is now deposited on the sample instead.
The orifice is widened after sputtering (here about \SI{130}{\um} diameter) and filled with redeposited material.
The EDS maps reveal a higher signal for \ce{Cu} at the orifice edges, probably due to pronounced sputtering, which may have removed the native or \ce{CO2}-plasma-induced \ce{Cu} oxide. 
Since \ce{N2} was mostly used as a gas with the nozzle as a cathode, \ce{N} is implanted at the orifice edge.
The lower row (Figure~\ref{fig:suppl:Nozzles}j) schematically shows the described deposition/removal of material on the nozzle depending on its polarity.

\subsection{Redeposition Effects with Nozzle as Cathode ($-$)}
\begin{figure*}[htb]
        
    \includegraphics[width=\linewidth]{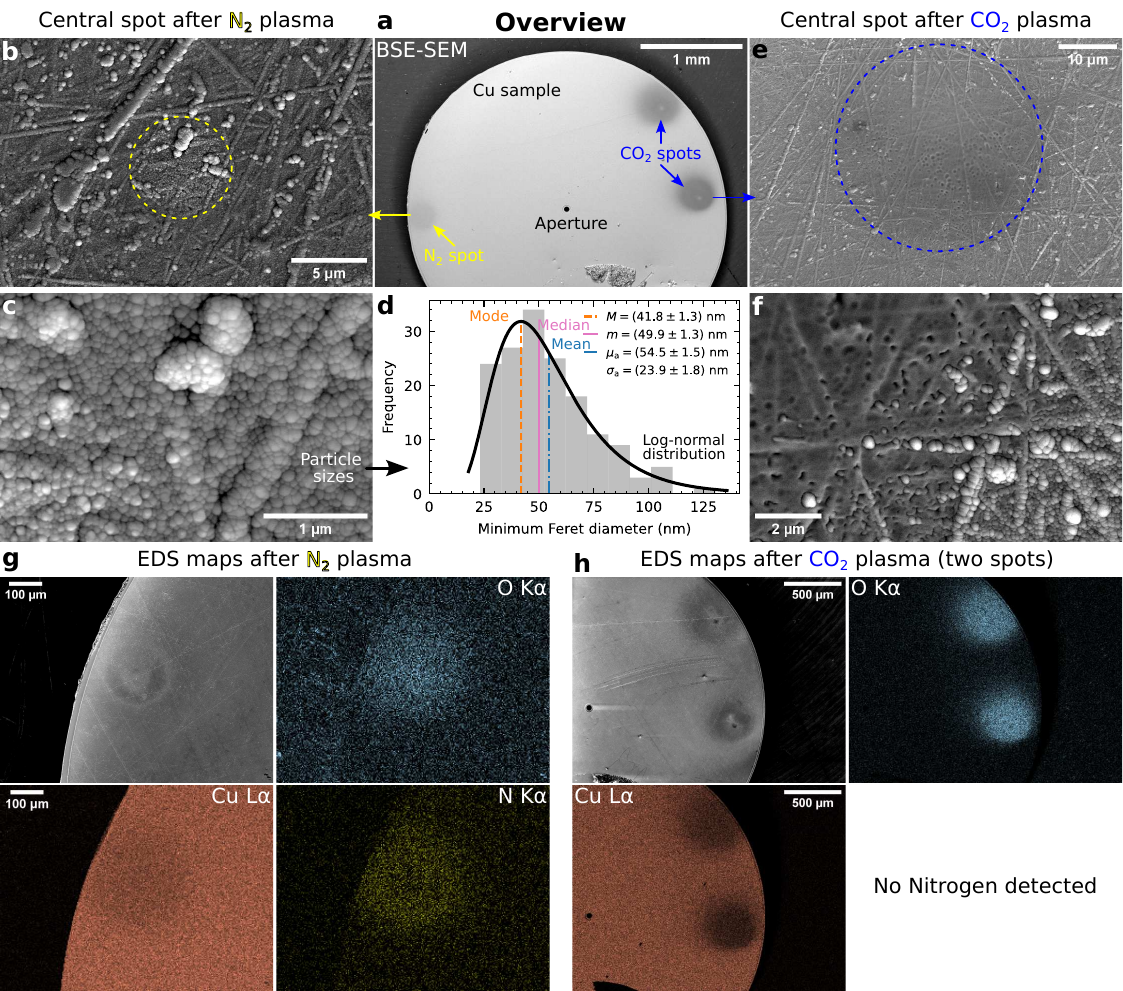}
    \caption{Comparison of \ce{N2} and \ce{CO2} plasma interactions on a \ce{Cu} surface with reversed polarity (nozzle as a cathode). \textbf{a}~Top-view BSE-SEM image of the \ce{Cu} foil used in the experiments. The left/right spot(s) result from \ce{N2}/\ce{CO2} plasma treatment. Magnified SE-SEM images of the central plasma regions for \ce{N2} (\textbf{b and c}) and \ce{CO2} (\textbf{e and f}) reveal grain stemming from redeposited nozzle material (here \ce{Cu} oxide). \textbf{d}~Grain size measurement from the SEM image in (c). \textbf{g}~EDS elemental maps reveal an \ce{O} signal besides \ce{N} after \ce{N2} plasma treatment, probably caused by \ce{Cu} oxide redeposition. \textbf{h}~EDS elemental maps reveal the oxidized spots diameters of about \SI{400}{\um} on the \ce{Cu} foil for \ce{CO2} plasma.}
    \label{fig:suppl:RedepositionN2}
        
\end{figure*}
Material from the sample is redeposited onto the nozzle by sputtering when the latter is used as anode.
\emph{Vice versa}, if the nozzle is used as a cathode, material from the nozzle is redeposited onto the sample~(Figure~\ref{fig:suppl:Nozzles}j).
Figure~\ref{fig:suppl:RedepositionN2} shows an analysis of this aspect after exposing a \ce{Cu} sample to \ce{N2} and \ce{CO2} plasmas with the nozzle used as a cathode.

The overview backscattered electron (BSE)-SEM image (taken at \SI{10}{\keV}), Figure~\ref{fig:suppl:RedepositionN2}a, shows reduced image intensity in the plasma-treated regions due to a locally reduced average atomic number $Z$.
Two spots on the right result from \ce{CO2} plasma treatment (marked in blue in Figure~\ref{fig:suppl:RedepositionN2}a), and the left one is from \ce{N2} plasma treatment (marked in yellow).
Higher magnification secondary electron (SE)-SEM images of the central plasma spots (marked with dashed circles in Figure~\ref{fig:suppl:RedepositionN2}b for \ce{N2} and Figure~\ref{fig:suppl:RedepositionN2}e for \ce{CO2}) reveal a finely-grained surface structure.
An approximate particle size of \SI[separate-uncertainty = true]{49.9(23.9)}{\nm} (arithmetic mean and standard deviation) was determined from Figure~\ref{fig:suppl:RedepositionN2}c by analyzing the minimum Feret diameters of the grains (Figure~\ref{fig:suppl:RedepositionN2}d). 
A similar surface structure is found for \ce{N2} (Figure~\ref{fig:suppl:RedepositionN2}c) and \ce{CO2} (Figure~\ref{fig:suppl:RedepositionN2}f).
The latter shows the transition region between the central plasma spot (flat) and the surrounding area (grains).
For \ce{CO2}, the central area is relatively flat with small pits, which is probably a result of ion$^{-}$ sputtering and longer plasma treatment duration than for \ce{N2}.\newline
Inspection of the elemental maps for \ce{CO2} reveals apparent oxidation of the plasma-treated region (Figure~\ref{fig:suppl:RedepositionN2}h). 
Similarly, the \ce{N2} plasma causes a local \ce{N} signal in EDS (Figure~\ref{fig:suppl:RedepositionN2}g), indicating \ce{N} implantation into the \ce{Cu} sample.
However, Figure~\ref{fig:suppl:RedepositionN2}g also shows an unexpected \ce{O} signal for \ce{N2} plasma treatment.
We suspect that this is caused by the redeposition of previously-oxidized \ce{Cu} that is present on the nozzle from earlier experiments (Figure~\ref{fig:suppl:Nozzles}j). 
The layer of redeposited (oxidized) \ce{Cu} is likely also the cause of the grain structure visible in Figure~\ref{fig:suppl:RedepositionN2}c and Figure~\ref{fig:suppl:RedepositionN2}f.
Independent of the gas used for plasma treatment, some material is sputtered from the nozzle when the latter is used as the cathode (negative polarity).
Material near the nozzle orifice is then primarily sputtered and redeposited onto the opposing sample.

\subsection{Details about the Experimental Setup}
\begin{figure*}[tb]
        
    \includegraphics[width=\linewidth]{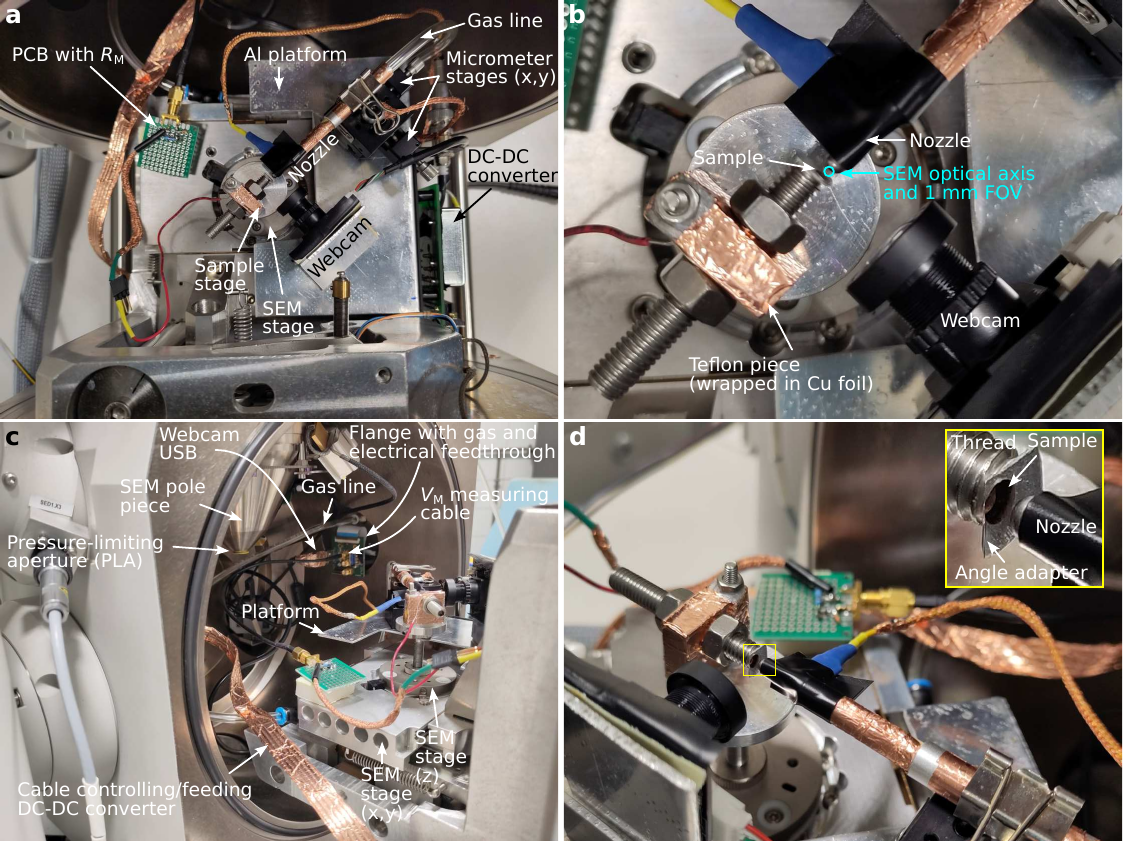}
    \caption{Photos of the experimental setup with opened SEM chamber. See the text for a detailed discussion.}
    \label{fig:suppl:SEM-Setup}
        
\end{figure*}
\begin{table*}[bt]
    \caption{Table of used components with their part number/stock-keeping unit (SKU).} 
    \label{tab:components}
    \begin{tabularx}{\textwidth}{l l l}
    \toprule
    Component & Manufacturer & Part number / SKU     \\
    \midrule
    Gas nozzle   & LenoxLaser  &  S-1/8-TUBE-CAL-20  \\
    Micrometer stages   & Thorlabs &  MS3/M  \\
    DC-DC converter (+2 kV)   & XP Power  &  CA20P  \\
    DC-DC converter (-1.25 kV)   & XP Power  &  CA12N  \\
    Power supply 1   &  Keysight  &  E36106B \\
    Power supply 2   &  RS PRO   &  IPS-3303 \\
    USB Webcam &    Arducam & B0205\\
    Flow meter   &  Alicat   &  M-200SCCM-D/5M \\
    Voltage measurement & Keithley & 2400 \\ 
    \ce{Cu}-aperture targets & Gilder Grids & GA50 \\
    \bottomrule
    \end{tabularx} 
\end{table*}

%
Figure~\ref{fig:suppl:SEM-Setup} shows photos of the assembled microplasma setup with opened SEM chamber.
The subfigures are explained in detail in the following, and the parts are listed in Table~\ref{tab:components}.
\begin{itemize}
    \item Figure~\ref{fig:suppl:SEM-Setup}a (top view): 
    Some components (gas nozzle, webcam, DC-DC converter) are mounted on an \ce{Al} platform to isolate them from the movements of the sample stage of the SEM. 
    The gas nozzle is fixed with a binder clip to two micrometer stages. 
    The high-voltage cable (starting at the BNC connection of the DC-DC converter, here visible below the gas line) is fixed close to the end of the nozzle and here hidden below the black insulating tape.
    The DC-DC converter was mounted on the side of the \ce{Al} platform to save space on the top. The sample is positioned in the center of the image opposite to the nozzle (Figures~\ref{fig:suppl:SEM-Setup}b and d).
    The printed circuit board (PCB, made according to the data sheet of the DC-DC converter, \url{https://www.xppower.com/portals/0/pdfs/SF_CA_Series.pdf}, with an output resistor of \SI{1}{\mega\ohm}) with the measurement resistor $R_\text{M}$ is fixed directly on the moveable part of the SEM stage. Its SMA connector and the cable go directly to the vacuum flange (Figure~\ref{fig:suppl:SEM-Setup}c).
    \item Figure~\ref{fig:suppl:SEM-Setup}b (top view): Close-up view of the nozzle-sample configuration.
    An adapter made of Teflon is used to isolate the sample from the SEM stage. 
    This choice was made to protect the SEM electronics from possible current bursts (e.g., arcing). The discharge current runs through the red cable, then through $R_\text{M}$ (see Figure~\ref{fig:suppl:SEM-Setup}a), and finally through the flange to the electronics outside the SEM chamber (not shown here).
    The \ce{Cu} foil on the top part of this Teflon piece is used to minimize charging.
    Since a \SI{500}{\um} diameter pressure-limiting aperture (PLA) was used, the field-of-view (FOV) for SEM imaging is significantly reduced.
    The actual visible area is around \SI{1}{\mm} and is exemplarily marked in the image with a circle. 
    The center of the circle is given by the optical axis of the SEM as manufacturer-calibrated to the $x=y=0$ position of the SEM stage.
    After setting the stage to $x=y=0$ without a mounted sample, the orifice of the nozzle is positioned as close as possible to the optical axis using the two micrometer stages. 
    The nozzle position is fixed after closing the SEM chamber.
    In case the nozzle is out of the SEMs FOV after pumping the chamber, its position has to be realigned after venting the chamber.
    When the nozzle is positioned within the SEMs FOV, the sample can be brought closer to/moved away from the nozzle using the SEM stage controls to change the gap distance.
    \item Figure~\ref{fig:suppl:SEM-Setup}c (side view): 
    The PLA is visible on the bottom of the SEM pole piece. 
    It reduces gas flow into the SEM column to keep it at higher vacuum levels compared to the plasma-gap region.
    A self-made flange with gas and electronic feedthroughs is mounted on one of the free chamber ports (here visible in the back).
    Notably, the gas line does a \SI{90}{\degree} bend in the feedthrough to account for x-ray safety.
    The cable of the DC-DC converter goes on the flange on the opposite side (not visible here, but in the top part of Figure~\ref{fig:suppl:SEM-Setup}d), where a DB9 connector is present. 
    This connector and the corresponding feedthrough are typically used for cooling/heating SEM stages provided by the microscope manufacturer.
    The pin layout was measured, and the shown custom cable with a DB9 connector was made to control the DC-DC converter.
    \item Figure~\ref{fig:suppl:SEM-Setup}d (side view): 
    The image from this angle reveals the sample stage made of an \enquote{angle adapter} to tilt the sample surface slightly toward the incident electron beam for analyses.
    Different angle adapters were ground with angles from \SIrange{5}{20}{\degree}. 
    These are fixed to the threaded metal rod with conductive \ce{Ag} paste.
    The sample used here is a \SI{3}{\mm} \ce{Cu} disc with a small aperture in the center (\SI{50}{\um} diameter, Gilder Grids GA50), which is glued onto the angle adapter with conductive \ce{Ag} paste.
\end{itemize}

The default vacuum system of the SEM (FEI Quanta 250 FEG) was used, consisting of a pre-vacuum rotary pump, a turbo molecular pump, and ion getter pumps (\textbf{IGPs}) for the electron-gun area.
Without gas flow, the SEM-chamber pressure was able to reach the \SI{4e-4}{\Pa} range after a few hours of pumping.
Especially, the residual air inside the gas line takes this time to get pumped through the nozzle orifice.
The gas lines were flushed with the process gas before experiments to reduce contamination with air.
With gas flow, the chamber pressure for a given gas flow rate depends on the gas type. 
The pressure is typically around \SI{2e-2}{\Pa} for gas flow rates of about \SI{5}{\sccm}, which is just below the threshold value of the microscope software for the high-vacuum mode (about \SI{3.3e-2}{\Pa}).
If the chamber pressure exceeds the threshold value, the gas flow rate must be reduced to allow for microscope and plasma operation in high-vacuum mode.\newline
It was observed that the use of \ce{Ar} leads to higher chamber pressures than for \ce{N2} or 
\ce{CO2} and, thus, \ce{Ar}-containing gas mixtures must be used more carefully.
The SEM chamber is mainly pumped by the turbo molecular pump, so a higher pumping speed for gases with smaller molecular weight is expected~\cite{beckerTurbomolecularPumpIts1966}, i.e., higher pumping speed for \ce{N2} (\SI{28}{\dalton}), followed by \ce{Ar} (\SI{40}{\dalton}), and finally \ce{CO2} (\SI{44}{\dalton}).
In practice, \ce{Ar} is probably less efficiently pumped than \ce{CO2} because the IGP of the electron column might contribute to the total pumping speed as well. 
As indicated by the microscope manufacturer in the microscope's manual, \enquote{the argon use should be minimized to a short time, because the
IGPs are not optimized for pumping of it at all.}, meaning that \ce{N2} and \ce{CO2} are likely to be more efficiently pumped by the IGP.

\clearpage
\bibliography{bibliography_bibtex}

\end{document}